%% file: main.tex
\documentclass{aa} %required by AA

\usepackage[english]{babel}
\usepackage[utf8x]{inputenc}
\usepackage[T1]{fontenc}
\usepackage{mathtools}
\usepackage[varg]{txfonts} %required by AA

\usepackage{threeparttable}
\usepackage{booktabs,caption}

\usepackage{graphicx}
\usepackage{epstopdf}
\usepackage{natbib}
\bibpunct{(}{)}{;}{a}{}{,} % to follow the A&A style 

\AppendGraphicsExtensions{.gif}

\usepackage{graphicx}
\usepackage[colorinlistoftodos]{todonotes}
\usepackage[colorlinks=true, allcolors=blue]{hyperref}
\usepackage{array}
\usepackage{booktabs}
\usepackage{pifont}

\begin{document}
%\tracingall %%this makes everything slower!!
\title{Tips and Tricks in linear imaging polarimetry of extended sources with FORS2 at the VLT
\thanks{Based on observations made with FORS2 at the VLT, ESO under programmes 60.A-9800(C) and 0101.D-0142(A)} %\fnmsep 
}
\author{S. Gonz\'alez-Gait\'an\inst{1}
\and A. M. Mour\~ao \inst{1,2}
\and F. Patat \inst{2}
\and J. P. Anderson \inst{3}
\and A. Cikota \inst{4}
\and K. Wiersema \inst{5}
\and A. B. Higgins \inst{6}
\and K. Silva \inst{3}
}
\institute{CENTRA-Centro de Astrof\'{\i}sica e Gravita\c{c}\~ao and Departamento de F\'{\i}sica, Instituto Superior T\'ecnico, Universidade de Lisboa, Avenida Rovisco Pais, 1049-001 Lisboa, Portugal 
\and ESO - European Southern Observatory, Karl-Schwarzschild-Str. 2, 85748, Garching b. München, Germany
\and European Southern Observatory, Alonso de C\'ordova 3107, Casilla 19, Santiago, Chile
\and Physics Division, Lawrence Berkeley National Laboratory, 1 Cyclotron Road, Berkeley, CA 94720, USA
\and Department of Physics, University of Warwick, Gibbet Hill Road, Coventry, CV4 7AL, UK
\and Department of Physics and Astronomy, University of Leicester, University Road, Leicester, LE1 7RH, UK
}
%\date{today}

%\abstractname
\abstract
{%Context: 
    Polarimetry is a very powerful tool to uncover various properties of astronomical objects that remain otherwise hidden in standard imaging or spectroscopic observations. While  common observations only measure the intensity of light, polarimetric measurements allow to distinguish and measure the two perpendicular components of the electric field associated with the incoming light. By doing so, it is possible to unveil asymmetries in supernova explosions, properties of intervening dust, characteristics of atmosphere of planets, among others.    
    However, the reliable measurement of the low polarization signal from astronomical sources requires a good control of spurious instrumental polarization induced by the various components of the optical system and the detector.
}
{%Aims:
    We perform a detailed multi-wavelength calibration study of the FORS2 instrument at the VLT operating in imaging polarimetric mode (IPOL) to characterize the spatial instrumental polarization that may affect the study of extended sources. 
}
{%Methods
    We use imaging polarimetry of a) high signal-to-noise blank fields $BVRI$ observations during full-moon, when the polarization is expected to be constant across the field-of-view and deviations originate from the instrument and b) a crowded star cluster in broad-band $RI$ and narrow-band $H_{\alpha}$ filters, where individual polarization values of each star across the field can be measured.
    }
{%Results:
    We find an instrumental polarization pattern that increases radially outwards from the optical axis of the instrument reaching up to 1.4\% at the edges, depending on the filter. Our results are well approximated by an elliptical paraboloid down to less than $\sim0.05\%$ accuracy,and $\sim0.02\%$ when using non-analytic fits. We present 2D maps to correct for this spurious instrumental polarization. We also give several tips and tricks to analyze polarimetric measurements of extended sources.
    }
{%Conclusions:
    FORS2 is a powerful instrument allowing to map the linear polarimetry of extended sources. We present and discuss a methodology to measure the polarization of such sources, and to correct for the spatial polarization induced in the optical system. This methodology could be applied to polarimetric measurements using other dual-beam polarimeters.}
\authorrunning{Gonz\'alez-Gait\'an, Mour\~ao et al.}
\titlerunning{FORS2 instrumental polarization}
\maketitle

\section{Introduction%: 
        }
        {\it 
        
        "Our scientific work in physics consists in asking questions about nature in the language that we possess and trying to get an answer from experiment by the means that are at our disposal"- \cite{heisenberg} }
        
    \input{sec1_introduction.tex}\label{sec:intro}

\section{Measurement of the polarization}\label{sec:theory}
        \input{sec2_polarization.tex}

\section{FORS2 instrumental characterization}\label{sec:fors2}
    The Focal Reducer and low dispersion Spectrograph FORS2 is a multi-mode optical instrument mounted on the Cassegrain focus of the UT1 telescope at ESO VLT.  It works in the wavelength range $330-1100$ nm. In imaging polarimetry mode (IPOL), one can measure the linear or linear/circular polarizations of the incident light, by employing a half-wavelength or quarter-wavelength plate, respectively. With the standard resolution (SR) collimator, 2x2 binning, the plate scale corresponds to 0.25"/pixel and to a field of view of 6.8'x 6.8' imaged onto two identical CCD detectors (CHIP1 and CHIP2). In this section, we describe the data used to characterize the instrumental polarization across the field of view of FORS2. A particular emphasis is given to the different steps in the analysis of imaging linear polarimetry of extended sources.
    
    \subsection{Data}

\input{sec_data}
     
     \subsection{Reduction steps}
     \input{sec_reduction}

     \subsection{The strip mask}\label{sec:strip}
     \input{sec_stripmask.tex}

     \subsection{Combining chips}\label{sec:combine}
     \input{sec_combine.tex}
     
     \subsection{Matching of the beams}\label{sec:match}
     \input{sec_matchbeams.tex}

     \subsection{Polarization flat}\label{sec:flat}
     \input{sec_flat.tex}

\section{FORS2 Instrumental polarization with moonlit sky}\label{sec:instpol_moon}
    \input{sec4_instpol_moon.tex}
    % to be done 

\section{FORS2 Instrumental polarization with a star cluster}\label{sec:instpol_cluster}
\input{sec_5_M30.tex}

\section{Stability of the instrumental polarization}\label{sec:discussion}
\input{sec_6_discussion.tex}

\section{Summary}\label{sec:sum}
We present a detailed study of the spatial instrumental polarization of FORS2 at VLT in imaging polarimetry mode for broad-band filters $BVRI$. We explore it also for the narrow-band filter $H_{\alpha}$. By studying moonlit sky blank fields and the stellar cluster M30, we find a radial instrumental pattern that is in good agreement with previous results for the decommissioned twin instrument FORS1 \citep{2006patat} in the broad-band filters. The radial behaviour is expected from the effect of the curved lenses of the collimator. The origin of some minor asymmetrical features, particularly in the $B$-band, remains unknown. We provide analytic equations to correct the $Q$ and $U$ Stokes parameters and the polarization in each band down to a $\sim$0.05\% level. We also release non-analytic INLA maps that enable a better correction of asymmetrical components down to $\sim0.02\%$. All reduction and analysis software, maps and results of this analysis are released.

We also present important tips and tricks to be taken into account in the reduction and analysis of polarimetric measurements of extended sources. We find for instance a variation of the mask strip positions in the CCD as a function of wavelength, and more importantly an increasing shift between $o$ and $e$-beam positions throughout the field that is well represented by a quadratic form. These effects come from the geometry of the Wollaston prism and the wavelength-dependence of its refractive indices. We also present a way to correct for flat-field in polarimetric observations by summing the images in all HWP angle positions of a given band.

The spatial polarization correction is vital in polarimetric studies of extended source such as galaxies, clusters, nebulae, among others. We have provided here the pathway for accurate corrections to an accuracy much below the typical estimated error in FORS2 at VLT. Although our initial studies suggest that the spatial instrumental linear polarization is stable, further studies of this kind at different observing conditions are encouraged to fully characterize its stability. Finally, the methodology presented in this study can be applied to polarimetric measurements using other dual-beam polarimeters. 

\section*{Acknowledgments}
%This work is based in calibration data obtained for the program P99A - "The first statistical study of multi-band optical polarimetry of supernovae host galaxies".    
We thank the anonymous referee for the valuable comments that improved the paper. We thank Jos\'e Lopes for some of the diagrams of this paper. The authors acknowledge  B. Leibundgut, G. Rupprecht and T. Szeifert for important clarifications regarding the characteristics and calibrations of FORS2 in IPOL mode, V. Ivanov for important discussions regarding data reduction, and S. Moehler for carefully reading the manuscript. A.M. thanks her colleagues at the Department of Physics, Instituto Superior T\'ecnico, namely P. Brogueira, I. Caba\c{c}o and C. Cruz for the very useful discussions about properties of crystals and G. Figueira for the tests with polarized light and a half wave plate. A.M. also thanks ESO for the hospitality and stimulating atmosphere and FCT \--\- Funda\c{c}\~ao para a Ci\^encia e Tecnologia for the financial support during her sabbatical leave at ESO. This work was supported by FCT under Project CRISP PTDC/FIS-AST-31546. The authors thankfully acknowledge the computer resources, technical expertise and assistance provided by CENTRA/IST. Computations were performed at the cluster “Baltasar-Sete-Sóis” and supported by the H2020 ERC Consolidator Grant "Matter and strong field gravity: New frontiers in Einstein's theory" grant agreement no. MaGRaTh-646597.

\bibliographystyle{aa}
%File with the references: references.bib
\bibliography{references.bib}

\appendix

\section{Rayleigh sky scattering}\label{ap:moon}
         \input{apB_Rayleigh_moon.tex}

\section{Fourier analysis}\label{ap:fourier}
         \input{ap_Fourier.tex}

\end{document}

%% file: sec1_introduction.tex
The polarization of light can carry relevant information about the emission and transmission processes occurring in astronomical objects, their surroundings and the material in the line of sight. 
Polarization is found in the light from nearby stars to stellar clusters, up to high-redshift galaxies,  and can be produced through emission mechanisms like synchrotron emission,  or indirectly through absorption or scattering in the interstellar medium (see \citealt{Hough06} for a review). From asymmetries in supernova explosions \citep[e.g.][]{Reilly17} to large scale magnetic fields in galaxies \citep[e.g.][]{Momose01}, polarimetry provides an extremely powerful tool to unveil their physical properties, otherwise hidden in common photometric or spectroscopic observations.

Despite the importance of polarimetric observations, carrying out polarization measurements can be a challenging task. 
Indeed, the measurement of the low degree of polarization of most astronomical sources [of order of few percent,  or even less than a percent] requires high signal to noise ratios (see below).  Moreover, spurious instrumental polarization might also induce artificial polarization signals. 
Today most large telescopes use the dual-beam configuration for polarimetric observations \citep[e.g.][]{Scarrott83}, which consists of a Wollaston prism to split the incoming light into two beams of perpendicular polarizations, a turnable retarder plate placed before the Wollaston prism to rotate the polarization plane, and a mask in the focal plane to avoid overlapping of the two polarization beams (see Figure~\ref{fig:Wollaston-schematic}).  

The present work is devoted to the study of the instrumental linear polarization induced by the dual-beam polarimeter of FORS2 - the FOcal Reducer and low dispersion Spectrograph\footnote{\url{https://www.eso.org/sci/facilities/paranal/instruments/fors/overview.html}} at the Europen Southern Observatory (ESO) Very Large Telescope (VLT). This study builds upon the work of \citet[][hereafter PR06]{2006patat}, which characterizes the instrumental polarization of the, now decommissioned, twin FORS1 instrument\footnote{FORS2 is largely identical to FORS1,  particularly its geometry and optical components, except for some modifications like grisms of higher resolution and coatings.} \citep{1998Appenzeller}. Although 
FORS2 has been widely used in polarimetric studies of point sources such as stars, supernovae, quasars and gamma-ray bursts, an increasing number of extended source observations demands for a full characterization of the spatial instrumental polarization. We provide thus here the methodology to analyze linear polarimetry of extended sources and to correct for the spurious instrumental pattern of the FORS2 dual-beam polarimeter.

This paper is divided as follows: in section~\ref{sec:theory} we present some basic concepts of polarization measurements with a dual-beam polarimeter, in section~\ref{sec:fors2} we present data and methods used for the instrumental characterization of the FORS2 polarimetric imaging mode across the field, in section~\ref{sec:instpol_moon} we present our results for the moonlit sky observations and in section~\ref{sec:instpol_cluster} the results for the stellar cluster observations. We examine possible changes of the instrumental linear polarization in section~\ref{sec:discussion}. We summarize our findings in section~\ref{sec:sum}.

%% file: sec2_polarization.tex
The estimation of the degree of polarization and the polarization angle of a partial polarized light beam
can be achieved using the methodology presented in PR06, which is based on the estimation of the four Stokes parameters $I,Q,U,V$,  where  I is the intensity of the source, Q,U describe the linear polarization and V the circular polarization. Here we consider that the incident light is partially linearly polarized, and so $V=0$\footnote{Although see the effect of cross-talk in section~\ref{sec:discussion}.}. It is possible to show that the Stokes parameters are related to the average intensities of the electric field components along two perpendicularly defined axes (see for instance \citealt{McMaster54,1960chandrasekhar, Born_Wolf_1980}). 

Following the notation of PR06, the polarization degree $P$ and the polarization angle $\chi$ are related to the Stokes parameters by:
\begin{equation}\label{eq:pol}
%\centering
P=\frac{\sqrt{Q^2 + U^2}}{I} \, \, ,
\end{equation}
%\noindent and
\begin{equation}\label{eq:polang}
\chi=\frac{1}{2} \arctan \frac{U}{Q} \, \, ,
\end{equation}

\noindent Hereafter, we use $Q\equiv Q/I$ and $U\equiv U/I$.

{  %fig 1
  \centering
\begin{figure}
\centering
\includegraphics[width=0.5\textwidth]{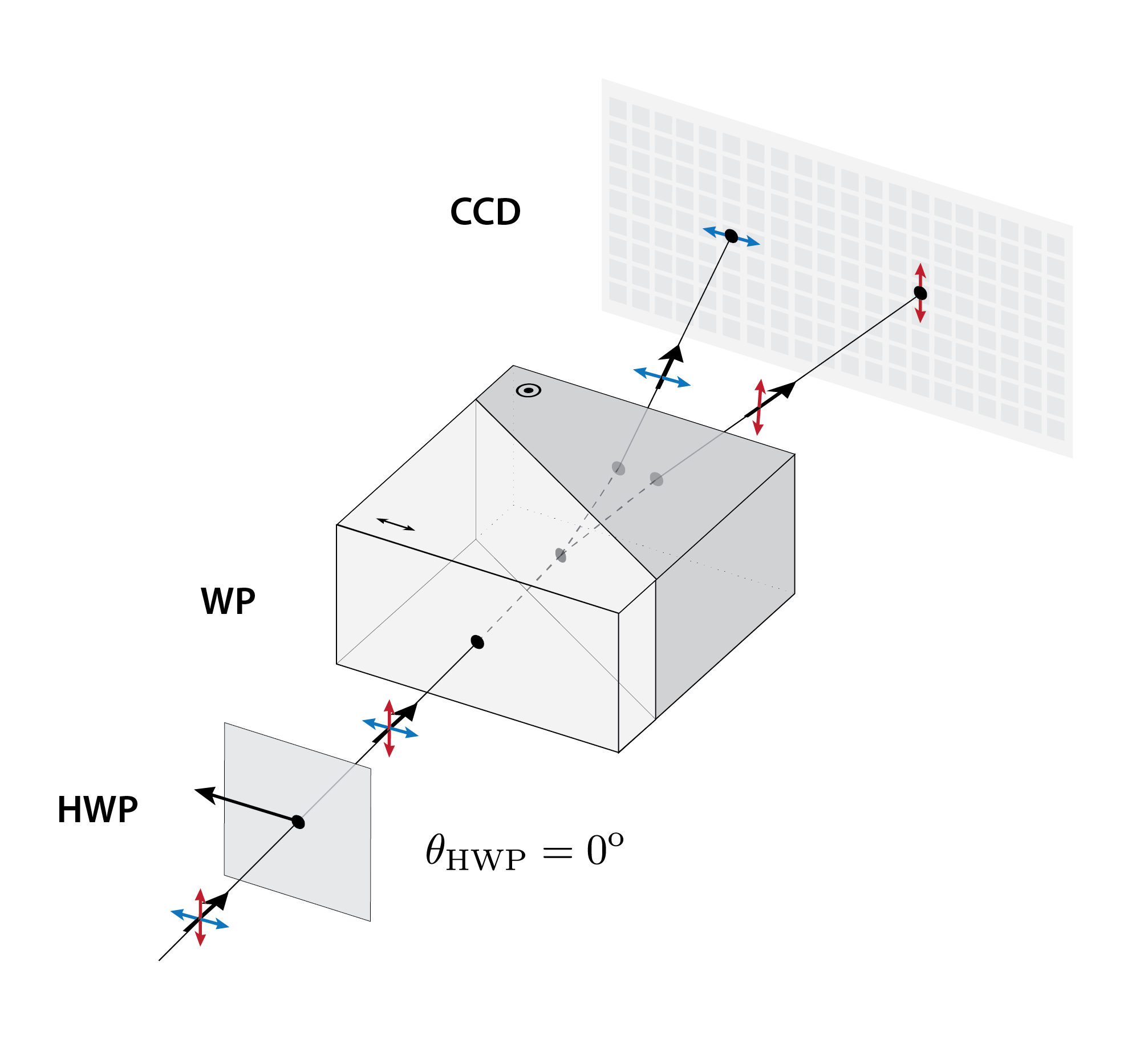}%1c:0.6
\caption{\label{fig:Wollaston-schematic} Schematic representation of a dual-beam polarimeter: an initial light ray with two polarization states (shown in blue and red) passes through a Half Wave Plate (HWP) and a Wollaston prism (WP) that splits it into two light beams with perpendicular polarization states. The optical axis in each  of the two crystals of the WP is also shown.} 
\end{figure}
~\\
 }

The polarization optics of dual-beam polarimeters, such as FORS2, allow us to measure simultaneously the intensities of two perpendicular components of the incident light beam, split by a Wollaston prism (WP, see Figure \ref{fig:Wollaston-schematic}). This measurement can be made at different position angles $\theta_i$ of the rotatable half-wave plate (HWP) that is placed before the WP. 
These two intensities are called ordinary $f_{O,i}$ and extraordinary $f_{E,i}$ beams\footnote{We note that according to FORS2 calibrations, when the HWP is at the zero position angle, the light from a point source is split into two beams in such a way that the component polarized along the meridian, called "ordinary beam", falls in the upper part of the $xy$ plane of the CCD whereas the beam with perpendicular polarization, called "extraordinary beam", reaches the CCD in the same column ($x$-position) but lower row ($y$-position).
The position angle of the HWP is measured relative to the meridian. Note that the terms "fast"/"slow" are also used to distinguish between the two out-coming perpendicularly polarized beams affected by lower/higher refractive indices in the WP and the HWP.}, and the total intensity $I$ of the source is related to $f_{O,i}$ and $f_{E,i}$ by $I=f_{O,i}+f_{E,i}$.

As discussed in PR06, the same field is observed at least at four HWP positions chosen at constant intervals of $\Delta \theta=\pi/8$ to minimize the errors. In this case, the normalized Stokes parameters can be obtained from 
\begin{equation}\label{eq:QU}
Q  = \frac{2}{N} \; \sum_{i=0}^{N-1} F_i \cos\left( \frac{\pi}{2}i \right) \qquad \textrm{and}\qquad 
U  = \frac{2}{N} \; \sum_{i=0}^{N-1} F_i \sin\left(\frac{\pi}{2}i\right)\,\, ,
\end{equation}

\noindent where each $F_i$, defined as the normalized flux difference at the position angle $\theta _i$, is given by:
\begin{equation}
\label{eq:FNi}
F_i \equiv \frac{f_{O,i} - f_{E,i}}{f_{O,i} + f_{E,i}} \, \,  . 
\end{equation}

\noindent It is worth mentioning that by adopting normalized flux differences, atmospheric variations are cancelled out. Thus, $F_i$ can be written as: 
\begin{equation}
\label{eq:FN_QU_P}
F_i = Q \cos 4\theta_i + U \sin 4\theta_i =
P\,\cos(4\theta_i - 2\chi)\, \, . 
\end{equation}

Finally, the absolute uncertainty in the  polarization degree, $ \sigma_P$, and the uncertainty in the polarization angle, $\sigma_{\chi}$, can be estimated  following the procedure of PR06, namely
\begin{equation}\label{eq:polerr}
\sigma_P =  \frac{1}{\sqrt{N/2}\times\textrm{SNR}}\qquad \textrm{and}\qquad
\sigma_{\chi} = \frac{\sigma_P}{2P} \, \, , 
\end{equation}
\noindent where SNR is the 
signal-to-noise ratio of the intensity, $I = f_O+f_E$. In appendix~\ref{ap:fourier}, we present an alternative way to estimate the error on the polarization degree and angle based on a Fourier analysis.

A correct estimation of the polarization of a given light source requires the control of any spurious instrumental polarization. The goal is to understand the possible origins of the instrumental polarization and to develop a methodology to correct for it. In the next chapters we describe the different steps needed for the reduction and analysis of polarimetric data in FORS2.

%% file: sec_data.tex
We acquired observatory technical time to characterize the instrumental
polarization across the field of FORS2. The data were
obtained during bright time (100\% lunar illumination at $\sim$18\textdegree\ from the Moon) on the 12th of March 2017 (between
1:00 and 2:00 UT). The moonlit sky
is expected to be highly polarized with a roughly constant value across the field of view. We targeted a blank field that is used for sky flats, $\alpha(\textrm{J2000})=$10h04m26.98s and $\delta(\textrm{J2000})=$-02\textdegree:19'00.26'',
ensuring that we obtained high signal, i.e. $>30000$ counts per
pixel. We observe the field with eight HWP position angles ($\theta_i=0$\textdegree, 22.5\textdegree, 45\textdegree,
67.5\textdegree, 90\textdegree, 112.5\textdegree, 135\textdegree\, and 157.5\textdegree) observed
in four filters, $b_{\textrm{HIGH}}$, $v_{\textrm{HIGH}}$, $R_{\textrm{SPECIAL}}$, $I_{\textrm{BESS}}$ (corresponding to ESO filter numbers 113, 114, 76 and 77), hereafter $BVRI$, with exposure times per HWP angle position of 200, 120, 100 and 135
seconds, respectively. 

Additionally, we used an independent way to measure the spatial instrumental
polarization with the help of individual mostly unpolarized stars across the field. For this, we also acquired data of the cluster M30 on a moonless night, at $\alpha(\textrm{J2000})=$21h40m22.12s and $\delta(\textrm{J2000})=$–23\textdegree10'47.5'', with four HWP angles
($\theta_i=0$\textdegree, 22.5\textdegree, 45\textdegree and 67.5\textdegree) in three different filters, $H_{\textrm{Alpha}}$, $R_{\textrm{SPECIAL}}$, $I_{\textrm{BESS}}$ (corresponding to filter numbers 83, 76 and 77), hereafter $H_{\alpha}RI$ under the program 0101.D-0142(A) (PI: Wiersema). The cluster was observed with 3 to 5 different small offsets in the $y$-direction to maximize the number of stars. Both datasets were taken with the red sensitive MIT CCD and with a 2$\times$2 binning. 

Finally, to test the stability of our results, we also used archival IPOL data of unpolarized and polarized standard star fields, namely: BD-12-5133, HDE-316232, Hiltner-652, Vela1, WD-0752-676, WD-1344+106, WD-1544-377, WD-1620-391, WD-2149+021 and WD-2359-434.

%% file: sec_reduction.tex
\begin{figure}%[t!]
\centering
  		\includegraphics[trim={0.cm 0.0cm 0.0cm 0.0cm },clip,width=0.4\textwidth]{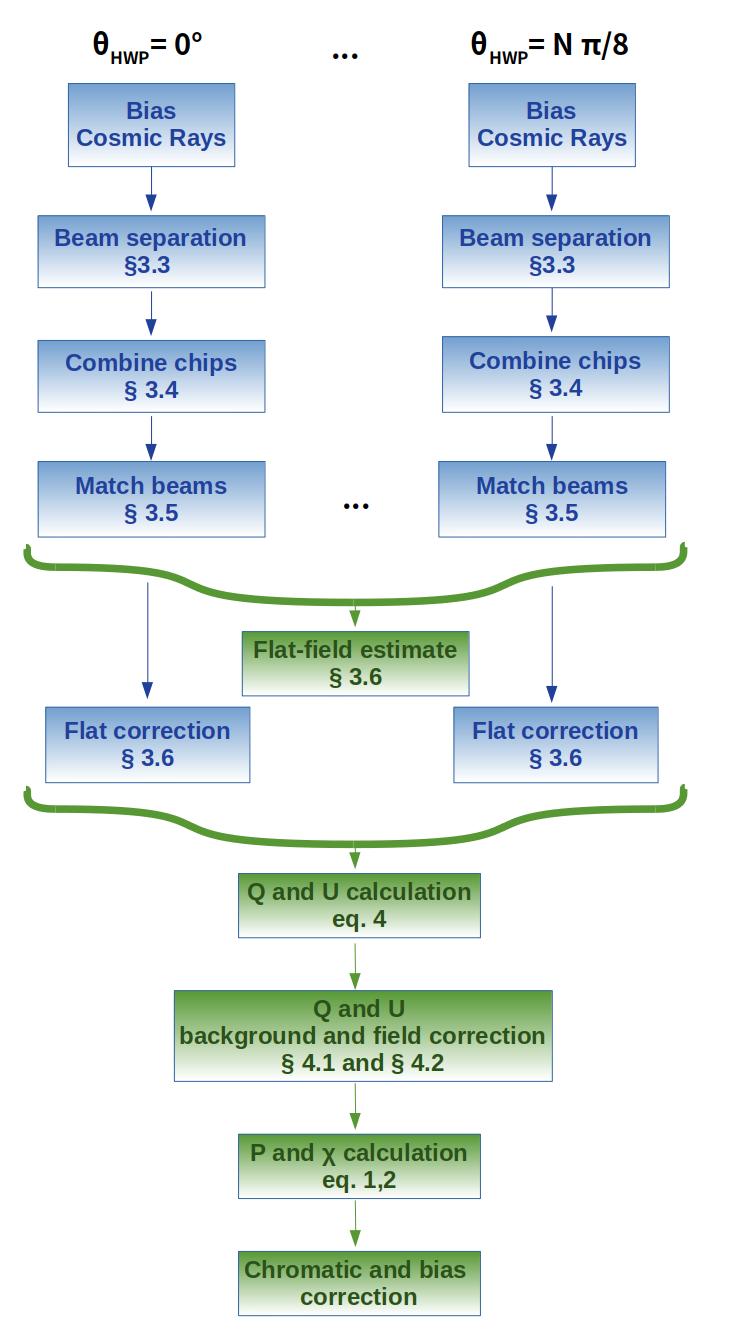}%1c:0.6
  		\caption{Flow chart of the different steps required for the reduction of imaging polarimetric data of extended sources with FORS2. Blue boxes indicate steps applied to every $\theta_{\textrm{HWP}}$ while green boxes require all HWP angles.}   
  		
  		\label{fig:flowchart}
\end{figure}

In figure~\ref{fig:flowchart}, we summarize the sequence of steps needed for the reduction of imaging polarimetric data of extended sources. Every raw image taken at each HWP angle is bias subtracted and corrected for cosmic rays \citep{VanDokkum01}. We then separate the $o$ and $e$-beams in each CCD using the strip mask (see section~\ref{sec:strip}) and combine both chips (see section~\ref{sec:combine}). We then match the $o$- and $e$-beam positions with the method described in section~\ref{sec:match}. We provide a way to apply a polarimetric "flat correction" in section~\ref{sec:flat}. The Stokes parameters need then to be corrected for the instrumental polarization values (section~\ref{sec:centralpol}) in order to obtain the polarization degree and angle (section~\ref{sec:pol}). Final steps include the chromatic correction of the zero angle of the HWP, which induces a non-negligible rotation in the polarization angle (see FORS2 manual and \citealt{Bagnulo09}), as well as the correction for the known polarization bias \citep{Serkowski58,Wardle74}, which tends to overestimate small values of $P$ with poor SNR. We use here the correction proposed by \citet{Plaszczynski14}. We make all of our software publicly available\footnote{See \url{https://github.com/gongsale/FORS2-INSTPOL/}}.

%REFLEX: ftp://ftp.eso.org/pub/dfs/pipelines/fors/fors-reflex-tutorial-1.6.pdf

%% file: sec_stripmask.tex
\begin{figure}%[t!]
\centering
	%\begin{minipage}[b]{.8\textwidth}
  		\includegraphics[width=0.50\textwidth]{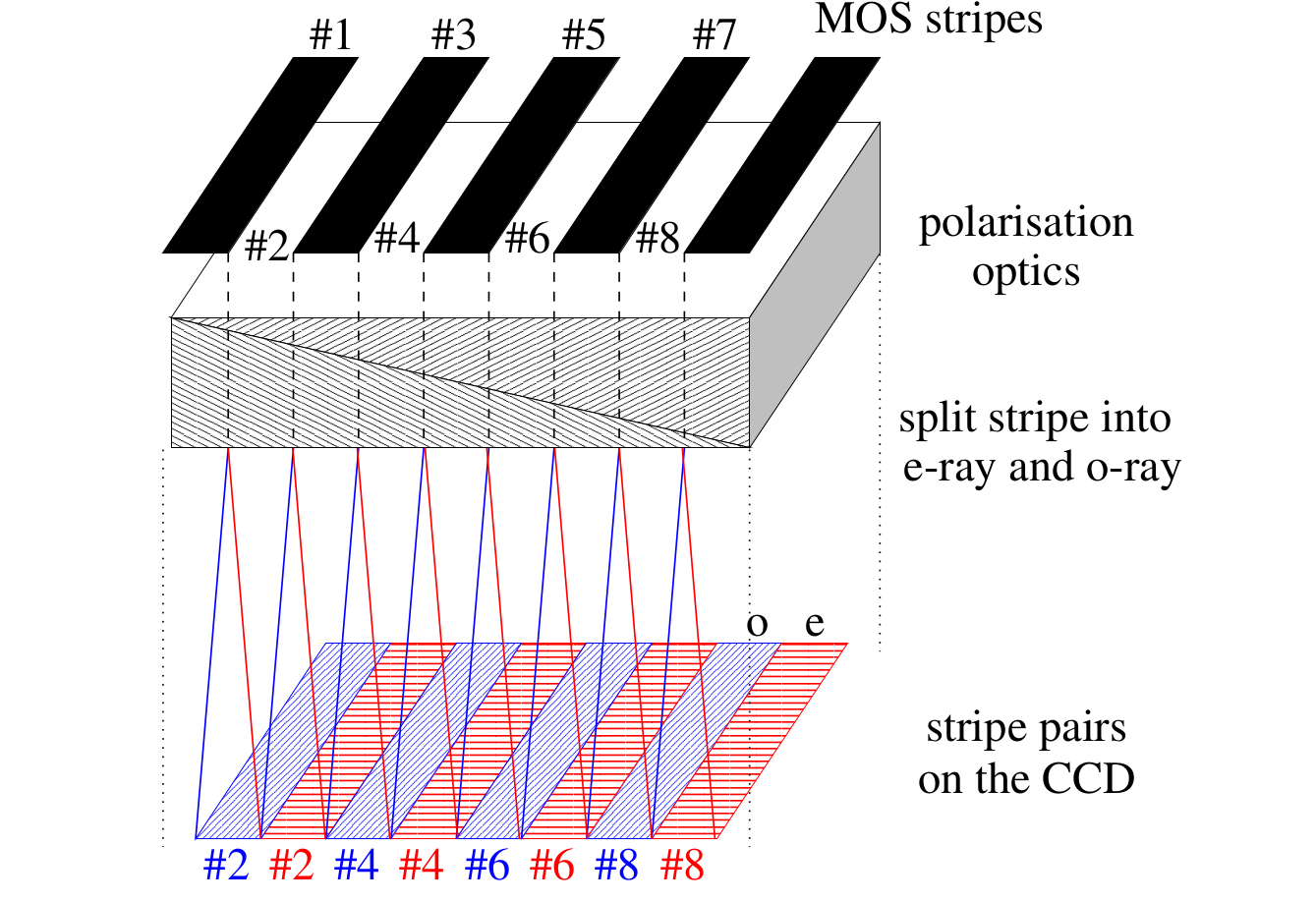}%1c: star,0.7
  		\caption{Schematic representation of the strip mask placed on the focal plane before the polarization optics and the resulting ordinary/extraordinary beam pairs on the CCD.	
  		\emph{Taken from the FORS2 manual.}}
  		\label{fig:mosstrips}
%	\end{minipage}
\end{figure}

In imaging polarimetry mode a strip mask is placed at the focal plane. This configuration avoids the superposition in the CCD of the two orthogonal polarized beams exiting from the birefrigent quartz WP. The strip mask is the odd numbered multi-object-spectroscopy (MOS) mask of 22" wide slits (see Figure~\ref{fig:mosstrips}). In principle, a full coverage of an extended source requires thus a minimum of two images displaced by 22", although three images are optimal as light gets lost at the mask edges due to diffraction effects. For observations of single point sources, the target is usually placed in the optical axis (located at the bottom of CHIP1) and requires no strip mask nor offset exposures.

The two polarization states of light split by the WP appear separated along the $y$-direction of the CCD in about 90 pixels.

We thus measure the strip $y$-positions for every single raw image of our moonlit and cluster observations by finding large abrupt changes in the light intensity gradient. We find no differences in the strip position for different HWP angles of a given filter. However, by summing all HWP images for a given filter, we do find variations in strip positions with wavelength of up to $\sim$3 pixels. We confirm this for several other fields of standard stars. This is expected as the refractive indices of the quartz are wavelength dependent, affecting the positioning of the strips on the CCD \footnote{Although the full analysis of the optical system is beyond the scope of the paper, in a simple treatment, the refractive indices of fuse quartz (SiO$_2$) change between 1.4663 at 440nm in $B$-band to 1.4539 at 768nm in $I$-band \citep{Malitson65} which leads to an outgoing angle difference of less than a degree and a corresponding vertical shift of up to 10 pixels at the edge the field.}. We report the final strip positions for all $BVRIH_{\alpha}$ filters in Table~\ref{table:strips}. Although variations of these positions are expected as the mask and/or instrument are removed and re-inserted, we find only $\lesssim2-3$ pixels deviations in other standard star fields. For more precise positions on a case-by-case, we recommend the gradient change approach that we follow here and that we release with our pipeline.

\begin{table}[h!t]
 \centering
 \caption{Strip $y$-positions in pixels for the two chips in all $BVRIH_{\alpha}$ filters. Strip labels (left column) are given in even pairs of \emph{o} and \emph{e}-beams (as in Fig.~\ref{fig:mosstrips})}. For each filter, left and right column indicate begin and end of each strip (where there is valid data). Strip positions increase from the bottom to the top of each chip, and CHIP1 is located above CHIP2. Note that strip number 8o of CHIP1 is cut and another part of it is found in CHIP2.\label{table:strips}
 \scalebox{0.7}{  %% not in 1c
  \begin{tabular}{c|cc|cc|cc|cc|cc}
    \hline
\hline
Strip Number & \multicolumn{2}{c}{$B$-band} & \multicolumn{2}{c}{$V$-band} & \multicolumn{2}{c}{$R$-band} & \multicolumn{2}{c}{$I$-band} & \multicolumn{2}{c}{$H_{\alpha}$-band} \\
    \hline
    \multicolumn{11}{c}{CHIP2} \\
    2e & 348 & 431 & 349 & 432 & 349 & 432 & 350 & 433 & 348& 434\\
    2o & 432 & 524 & 433 & 523 & 433 & 521 & 434 & 521 & 435& 523\\
    4e & 530 & 612 & 531 & 612 & 531 & 612 & 532 & 613 & 532& 614\\
    4o & 623 & 705 & 622 & 703 & 621 & 702 & 620 & 702 & 621& 704\\
    6e & 711 & 795 & 712 & 796 & 712 & 796 & 713 & 797 & 712& 798\\
    6o & 804 & 889 & 803 & 887 & 802 & 886 & 802 & 886 & 802& 888\\
    8e & 891 & 973 & 892 & 974 & 892 & 974 & 893 & 975 & 893& 975\\
    8o & 985 & 1027& 983 & 1027& 982 & 1027& 982 & 1027 &983& 1027\\
    %9 & & & & & & & & \\
    \hline
    \multicolumn{11}{c}{CHIP1} \\
    8o  & 5   & 27  & 5   & 25  & 5   & 24  & 5   & 24  & 5  & 25 \\
    10e  & 32  & 113 & 32  & 114 & 32  & 114 & 33  & 115 & 33 & 116\\
    10o  & 126 & 208 & 124 & 206 & 123 & 205 & 123 & 205 & 123& 207\\
    12e  & 212 & 295 & 212 & 296 & 213 & 296 & 213 & 297 & 213& 297\\
    12o  & 307 & 390 & 305 & 389 & 304 & 388 & 304 & 387 & 304& 389\\
    14e  & 394 & 476 & 395 & 477 & 395 & 477 & 396 & 478 & 395& 478\\
    14o  & 490 & 572 & 488 & 570 & 487 & 569 & 487 & 569 & 487& 571\\
    16e  & 574 & 656 & 575 & 657 & 575 & 657 & 576 & 658 & 576& 658\\
    16o  & 671 & 753 & 669 & 752 & 668 & 750 & 668 & 750 & 669& 752\\
    18e & 755 & 838 & 756 & 838 & 756 & 838 & 757 & 839 & 756& 841\\
    18o & 853 & 936 & 851 & 934 & 850 & 933 & 850 & 933 & 850& 936\\
    %12 & 937 & & & & & & & \\
    \end{tabular}
}
\end{table}

%% file: sec_combine.tex
Since the FORS2 data consists of mosaics of two images from two CCDs or chips, we need to combine both to study extended sources. We briefly highlight here the procedure. To begin, it is important to note that besides the CCD gap in $y$ of $y_{\textrm{GAP}}=480\mu m$ (32pix), there is also a shift in $x$ of $x_{\textrm{GAP}}=30\mu m$ (2pix) and also a slight angle ($\theta_R=0.083$\textdegree) between both chips. This requires both, a rotation and a translation, that we perform on the second, bottom, CHIP2. 

If $x_2$ and $y_2$\footnote{We note that there is a pre- and over-scan region for both chips of FORS2 of 5 pixels in $y$ which we previously remove.} are the coordinates in CHIP2 in $\mu m$, using a pixel size of 15$\mu m$, the simultaneous rotation and translation consist of:

\begin{eqnarray}
x_2' = x_2\cos(\theta_R) - y_2\sin(\theta_R)-x_{\textrm{GAP}}\\ \nonumber
y_2' = x_2\sin(\theta_R) + y_2\cos(\theta_R)-y_{\textrm{GAP}}, 
\end{eqnarray}
where $x_2'$ and $y_2'$ are the new positions of CHIP2. The flux at those new positions are found via cubic 2D interpolation. We have checked that the interpolation conserves the input signal well within the estimated flux uncertainties. After the combination of the two chips, the new reference point ($x=0,y=0$) lies at the bottom left pixel of the combined image. The typical position of the optical axis where a point source is usually observed lies at $x_0=1226$ pix and $y_0=1025$ pix of the combined image.

%% file: sec_matchbeams.tex
After obtaining the two combined images of the $o$- and $e$-beams, one needs to accurately match the coordinates of the ordinary and extraordinary beams corresponding to the same sky positions. This is required in order measure the corresponding polarization in each position. In principle, the position in $x$ should be the same, while the shift in the $y$-position is expected to be of the order of the width of each strip, i.e. around 90 pixels. Given that the refractive index of the WP changes with wavelength, one might expect this shift in $y$ to change with filter as well. 

\begin{figure*}%[t!]
\centering
  		\includegraphics[width=\textwidth]{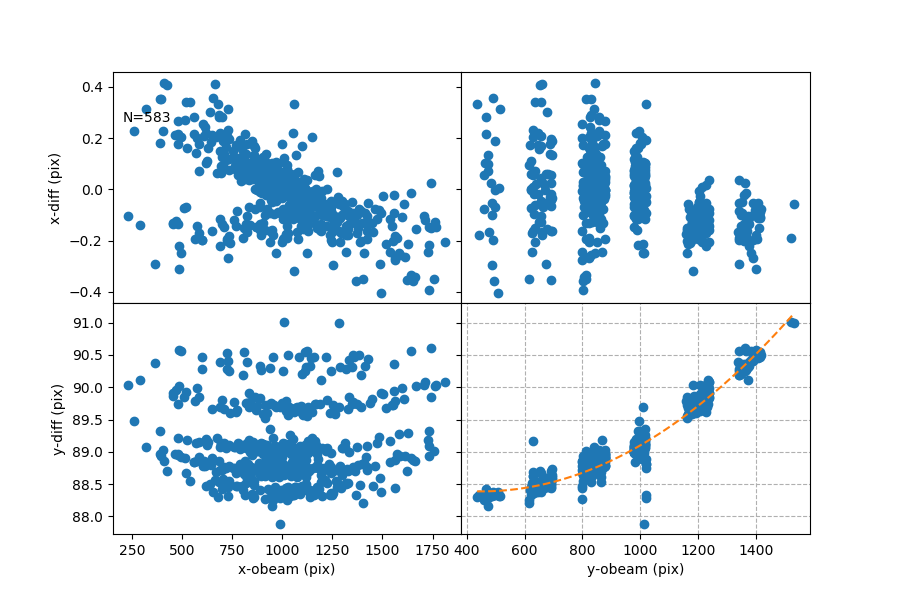}
  		\caption{Difference in $o$- and $e$-beam position, $x_{\textrm{diff}} = x_O-x_E$ (upper) and $y_{\textrm{diff}} = y_O -y_E$ (bottom), for matched stars of ordinary and extraordinary beams as a function of $o$-beam $x_O$ (left) and $y_O$ (right) location in the combined image. The results correspond to 583 stars of a M30 $I$-band image at $\theta_{\textrm{HWP}}=22.5$\textdegree. Best quadratic fit for $y_{\textrm{diff}}$ vs $y$ is shown in orange.}
  		\label{fig:quadpars}
\end{figure*}	

In the case of point sources, one may directly use the source to find the shift between the $o$ and $e$-beams. To determine this for extended sources, we take advantage of the field stars by matching the brightest stars found in each of the two beam images of all of our M30 cluster and Vela images. We use {\sc daofind} \citep{Stetson87} for {\sc python} and take the closest match in $x$ and $y$ positions for each star,
ensuring that they do not exceed $\Delta x < 10$ pix and $\Delta y < 150$ pix to avoid incorrect matches. We show an example in Figure~\ref{fig:quadpars} where we can see that, as expected, $\Delta x < 1$ pixel, whereas the difference in $\Delta y \sim 90$ pixels is not constant across the CCD but increases by up to $\sim3-7$ pixels, depending on the filter, as one moves towards the top of the combined image. This comes from the fact that although the angular separation between the beams (arising from the difference in refractive indices) is constant, its corresponding spatial projection in the vertical plane of the CCD varies. We find that this difference is very well approximated by a quadratic form:

\begin{equation}\label{eq:match}
    \Delta y [\textrm{pix}]= (y_O-y_E) = a+b\,(y_O-450)^2,
\end{equation}

\noindent where $a$ and  $b$ are free parameters in the fit and $y_O$ and $y_E$ are the pixel positions of the ordinary and extraordinary beams, respectively. This behaviour is not symmetric about the center of the field of view but increases from bottom to top forming a single quadratic form. 
We do not see any variation of the parameters with respect to HWP angle but we find an evident shift with filter coming from the wavelength dependence of the refractive indices of the WP. The final parameters for each filter are given in Table~\ref{table:quadpars}. We find these to be very robust ($\Delta y<$1pix) across different analyzed fields and epochs. Finally, we also note that there is a slight trend of $\Delta y$ and $\Delta x$  with $x$-position, perhaps due to geometrical effects in the optics, but this is always below 0.5 pixel which is within the error in the pixel position and much less than the typical PSF of the image and can thus be ignored. All of our $e$-beam image positions are matched to the corresponding $o$-beam through the relation given in eq. \ref{eq:match}, so that later fluxes in each beam can be compared and polarization analysis carried out.

\begin{table}[h!t]
 \centering
 \caption{Quadratic parameters to match $y$-positions of ordinary and extraordinary beams for filters $BVRIH_{\alpha}$. Errors quoted correspond to the standard deviation from images in all HWP angles.}\label{table:quadpars}
 \scalebox{0.7}{
 \begin{tabular}{c|ccccc}
    \hline
\hline  
Parameter & $B$-band & $V$-band & $R$-band & $I$-band & $H_{\alpha}$  \\
\hline
$a$ & 92.56 $\pm$ 0.03&  90.56 $\pm$ 0.03& 89.53 $\pm$ 0.01 & 88.39 $\pm$ 0.01 & 89.46 $\pm$ 0.01\\ 
$b\,[10^{-6}]$&  2.42 $\pm$ 0.03& 2.30 $\pm$ 0.01& 2.31 $\pm$ 0.06 & 2.33 $\pm$ 0.01 & 2.31 $\pm$ 0.04  \\
    \end{tabular}
    }
\end{table}

%% file: sec_flat.tex
In dual-beam polarimetry, the flat-fielding correction is a challenging task. Since the beam is split after optical components like the focal mask, the collimator and the HWP, flats need to be taken with these elements in the light path. However, sources like internal flat screens and twilight sky may introduce polarization that is difficult to eliminate. In principle, one can mitigate this polarizing effect by obtaining dome screen flats with a continuously rotating HWP, so that the rotation is faster than the exposure time \citep[see e.g.][]{Wiersema18}. As this is not possible for FORS2, we construct depolarized flats by summing all HWP angles for the combined and matched $o$ and $e$-beams:

\begin{equation}
f_O = \sum_{i}^{N} f_{O,i} \quad\textrm{and}\quad f_E = \sum_{i}^{N} f_{E,i},
\end{equation}

\noindent where $i$ corresponds to the index the HWP position angle. 
 
\begin{figure*}
\centering
\includegraphics[trim={5.9cm 0 4.0cm 0 },clip,width=0.49\textwidth]{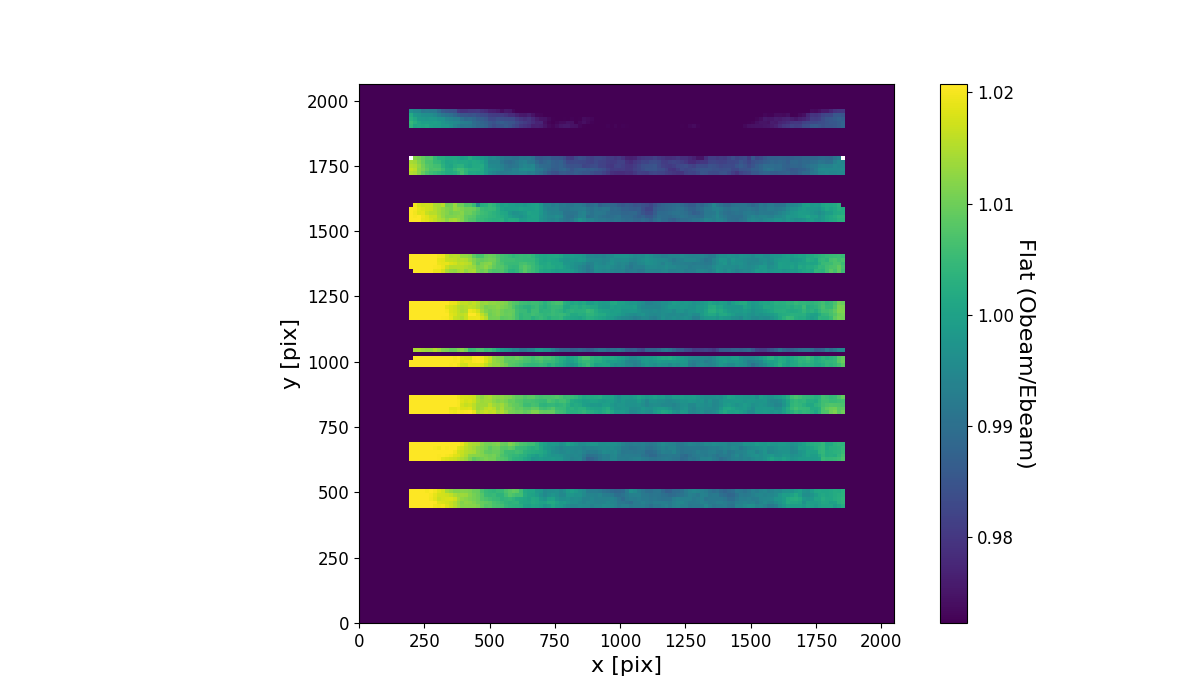}
\includegraphics[trim={5.9cm 0 4.0cm 0 },clip,width=0.49\textwidth]{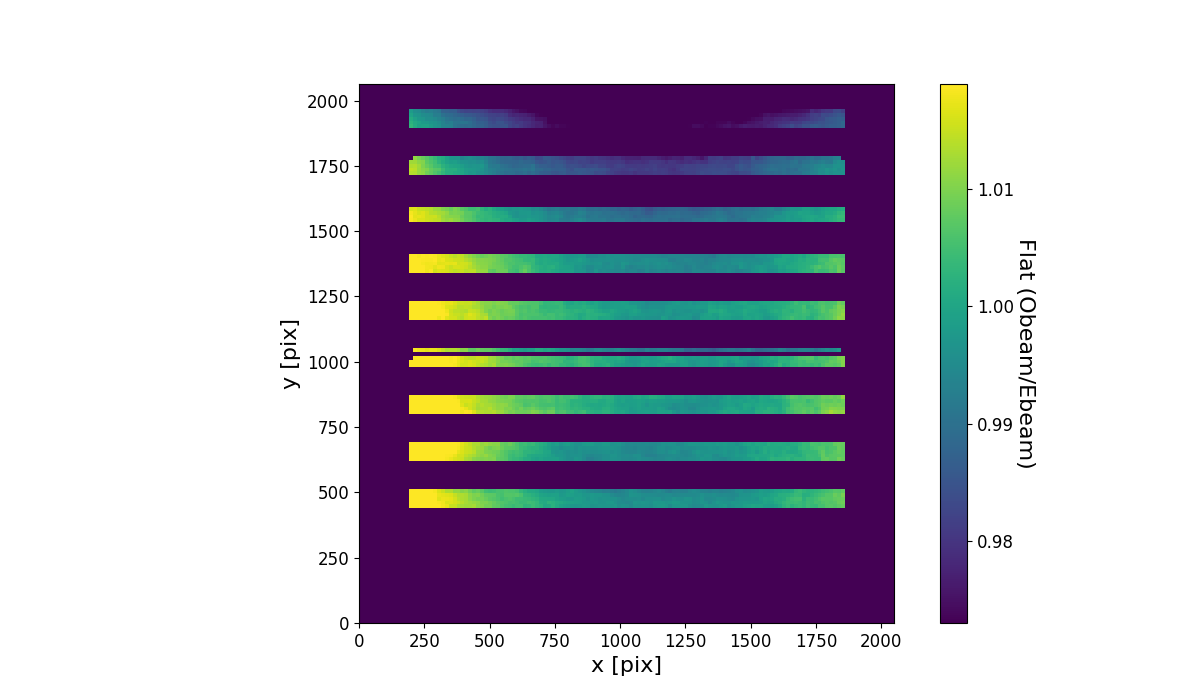}
\caption{\label{fig:flat} Binned (30$\times$30 pixel) polarization flat: sum of ordinary beam in 8 HWP angles divided by the sum of extraordinary beam in 8 HWP angles for a moonlit blank sky field in $V$-band (left) and $R$-band (right).}
\end{figure*}

In the ideal case, both $f_O$ and $f_E$ flats should be equal under uniform illumination of the field. We thus use here our blank field observations during full moon which, for a given filter, have essentially constant illumination. In Figure~\ref{fig:flat}, we show the polarization flat-field, defined as the ratio of the summed angles in $o$- to $e$-beam defined above: flat=$f_O/f_E$. Differences in the corresponding $o$ and the $e$-beams might indicate deviations of the WP from uniformity. Thus, this flat can be used to correct the beam intensities before calculating the flux differences (eq.~\ref{eq:FNi}) and the $Q$ and $U$ Stokes parameters (eq.~\ref{eq:QU}).

We see that the flat-field may induce a 2\% correction at the edge of the field. However, when we apply this flat correction before measuring the polarization, we find no significant difference in the maps of polarization degree and angle (see next section) due, in part, to the redundancy of multiple $N=8$ HWP angles. However, by doing a careful Fourier analysis in appendix~\ref{ap:fourier}, we find that the flat correction reduces the errors in the polarization degree arising from secondary harmonics down to $\lesssim0.4$\%. Besides this small effect, flat-fielding can also help to clip bad pixels from a background measurement. Also, being taken at the same instrument position, such a flat is useful to correct for fringing, an effect that is generally small for FORS2, but could become more important at longer wavelengths, e.g. in filter $z$, or with different detectors like the blue-sensitive E2V CCD, which is known to present more fringing.
We therefore advise to do the flat correction when dealing with extended sources.

%% file: sec4_instpol_moon.tex
The optical components of the telescope system and of the detector may induce instrumental polarization, as already highlighted for FORS1 (PR06). Disentangling the contribution of each component of the optical system for the total spurious instrumental polarization is a difficult task to simulate. In this section, we report on experimental tests and observations that can bring out the characteristics  of the spurious instrumental polarization in the FoV of FORS2. 

\subsection{Background polarization}\label{sec:centralpol}

One way to estimate large scale instrumental polarization is by measuring the moonlit sky polarization during full moon. The small field of view of the FORS2 ensures a roughly constant polarization of the Moon light along the image. The moonlit sky polarization depends on the separation of the observed target to the Moon and on the observing date and time. Observations of small fields \citep{Wolstencroft73} and simulations of single Rayleigh scattering predict less than 0.05\% polarization variance in sky polarization within the FoV of the instrument. For this reason, larger deviations from a constant polarization in the FoV are expected to be independent of the Moon and generated by the spurious instrumental polarization.

\begin{figure*}
\centering
\includegraphics[width=0.80\textwidth]{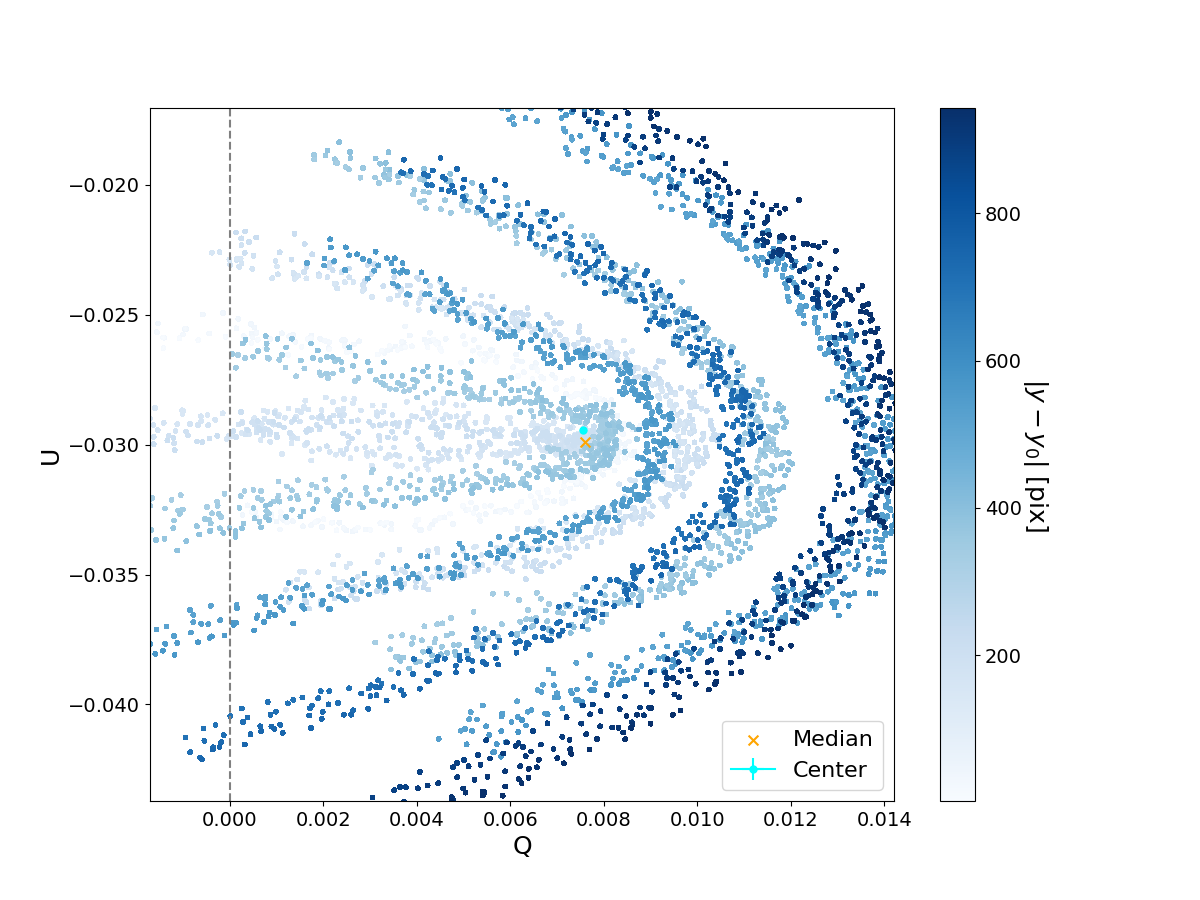}
\caption{\label{fig:QUplane} $Q$-$U$ plane for moonlit sky polarization in $R$-band. Each point corresponds to a box of $30\times30$ pixels and the $y$-position with respect to the center is colored in different blue shades. The median value of $Q$ and $U$ is shown as an orange cross while the value at the optical axis ($x_0=1226,y_0=1025$pix) of the instrument is shown in cyan. Dashed lines indicate $Q=0$. Typical errors on $Q$ and $U$ are of the order of $10^{-5}$ (see Table~\ref{table:error}).}
\end{figure*}

Throughout this section, we study the field polarization. We analyze both, pixel by pixel, and also in binned boxes of $30\times30$ pixels in $f_{O,i}$, $f_{E,i}$ images to increase the S/N by measuring the $2\sigma$-clipped mean intensity of each box. We always find that the results obtained pixel by pixel are consistent with the binned values, also for different box sizes and clipping.

To estimate the instrumental polarization, we follow the methodology of section~\ref{sec:fors2}: all of the 16 images (2 chips in 8 HWP angles) of a given filter are first separated into $o$ and $e$-beams, chip-combined and flat-field corrected; then we use the eight HWP angles to measure $Q$ and $U$ Stokes parameters with eq.~\ref{eq:QU}. 
In fig.~\ref{fig:QUplane}, we show the $Q$-$U$ plane in $R$-band. Clearly, there is an overall shift of all points with respect to zero that is best represented by the median background value of all points, at $Q_R\simeq0.8$\% and $U_R\simeq-2.9$\%, which is very close the value at the center of the FoV. This corresponds (from eq.~\ref{eq:pol}) to a polarization of $P_R=3.0$\%. 

We note that using a simple Rayleigh scattering model of the Moon located at $\sim17$\textdegree\, from our blank target field (see Appendix~\ref{ap:moon}) predicts a background value of $P_B=3.5$\%. All background values for our $BVRI$ images are shown in Table~\ref{table:cenQU}. We analyze fields with different background polarizations in section~\ref{sec:discussion}.

\begin{table}[h!t]
\centering
 \caption{Background $Q_B$, $U_B$, $P_B$ and $\chi_B$ values of moonlit sky observations for $BVRI$ filters. The polarization angle has been corrected for the chromatism of the HWP (see FORS2 manual\protect\footnotemark). }\label{table:cenQU}
 \begin{tabular}{c|cccc}
    \hline
\hline  
Filter & $Q_B$[\%] & $U_B$[\%] & $P_B$[\%] & $\chi_B$[\textdegree] \\
\hline
$B$-band  & 1.12$\pm$0.11 & -2.17$\pm$0.11 & 2.44$\pm$0.11 & 148.4$\pm$0.2\\ 
$V$-band & 1.63$\pm$0.12 & -3.42$\pm$0.12 & 3.79$\pm$0.12 & 147.7$\pm$0.1\\
$R$-band & 0.77$\pm$0.12 & -2.93$\pm$0.13 & 3.03$\pm$0.13 & 142.3$\pm$1.9\\
$I$-band & 0.18$\pm$0.12 & -2.23$\pm$0.14 & 2.23$\pm$0.12 & 137.3$\pm$2.8\\
    \end{tabular}
\end{table}

To analyze the spatial characteristics of the instrumental polarization we must therefore correct for these background values before calculating the polarization degree and angle along the FoV using eq.~\ref{eq:pol}. In principle this background polarization corresponds to the central value (i.e. at the telescope optical axis), but in fields with sources in the center, the correction with the median is more appropriate. We thus do a vectorial correction of the observed polarization:  

\begin{eqnarray}
    Q_{\textrm{corr}} & = & Q - Q_B \\
    U_{\textrm{corr}} & = & U - U_B\ , \nonumber
\end{eqnarray}
where $Q_B$ and $U_B$ are the background median values.
\footnotetext{\url{http://www.eso.org/sci/facilities/paranal/instruments/fors/inst/pola.html}}

Alternatively, by substituting this into eq.~\ref{eq:pol},the measured polarization parameters $P,\chi$, corrected for the background values, $P_B,\chi_B$, are given by:

\begin{eqnarray}
    P_{\textrm{corr}} & = & \sqrt{P^2+P_B^2-2PP_B\cos{(\chi-\chi_B)}} \\
    \chi_{\textrm{corr}} & = & \arctan{\left[\frac{P\sin{\chi}-P_B\sin{\chi_B}}{P\cos{\chi}-P_B\cos{\chi_B}}\right]} \nonumber
\end{eqnarray}

\begin{figure*}
\includegraphics[trim={1.4cm 1.5cm 1.2cm 1.5cm },clip,width=0.5\textwidth]{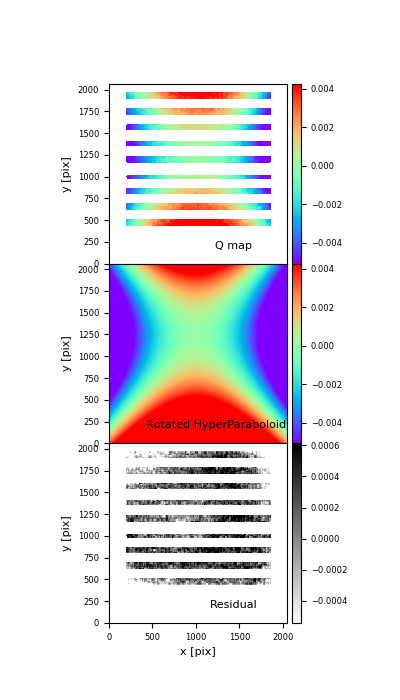}
\includegraphics[trim={1.4cm 1.5cm 1.2cm 1.5cm },clip,width=0.5\textwidth]{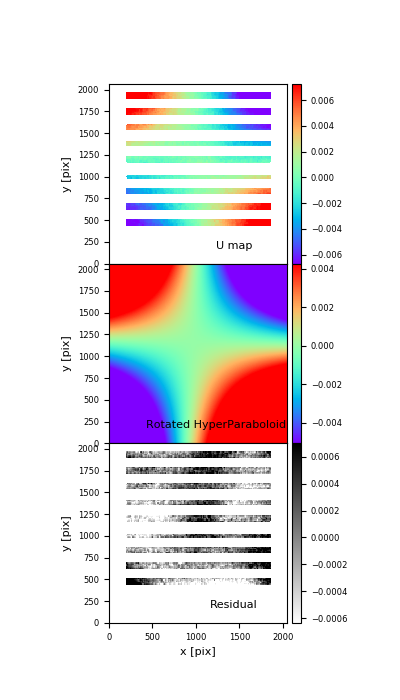}
\caption{\label{fig:QU} \emph{Top}: Binned $Q$ (left) and $U$ (right) maps in $B$-band after correction for the background value. \emph{Middle}: Hyperbolic paraboloid fit using eq.~\ref{eq:hypparab} and parameters in Table~\ref{table:hypparab}. \emph{Bottom}: Residual of observed map minus the fit.}
\end{figure*}

However, even before correcting for the background polarization, we can already see in the uncorrected $Q$-$U$ plane of Figure~\ref{fig:QUplane} that there is a net increase in the $Q$ and $U$ parameters as one moves away from the optical axis. In fact, one may appreciate the $y$-strip map structure that produces the empty spaces in the $Q$-$U$ plane. We show the $Q$ and $U$ maps \emph{after} correcting for the background values in Figure~\ref{fig:QU} for the $B$-band. Both $Q$ and $U$ patterns are well approximated by a hyperbolic paraboloid (a "saddle") of the form:

\begin{eqnarray}\label{eq:hypparab}
     Q(x,y),U(x,y) &=& \frac{[y'(x,y)]^2}{b^2} - \frac{[x'(x,y)]^2}{a^2} \\ \nonumber
     &=&\frac{[(x-x_0)\sin\theta+(y-y_0)\cos\theta]^2}{b^2} \\ \nonumber &-&\frac{[(x-x_0)\cos\theta-(y-y_0)\sin\theta]^2}{a^2} \nonumber
\end{eqnarray}

The paraboloid is described by the free $a$ and $b$ parameters (different for $Q$ and $U$), whereas $x_0$ and $y_0$ represent the central position where $Q,U=0$, and $\theta$ determines the rotation of the paraboloid in the $xy$ plane, so that $x',y'\xrightarrow{\theta} x,y$. Such a 2D fit and residual map are also shown in Figure~\ref{fig:QU}. All residuals are below 0.06\% for $Q$ and $U$. We present the resulting fit parameters for $BVRI$ in Table~\ref{table:hypparab}. The analytic maps can be used to correct the spatial instrumental polarization.

\begin{table*}[h!t]
\centering
 \scalebox{1.00}{
\begin{threeparttable}
 \caption{Parameters for the rotated hyperbolic paraboloid of eq.~\ref{eq:hypparab} for $Q$ and $U$ vs $x$ and $y$ positions (pixels) for $BVRIH_{\alpha}$ filters. The origin is taken at the bottom left of the combined image}.\label{table:hypparab}.
 \begin{tabular}{c|cccccc}
    \hline
\hline
& \multicolumn{6}{c}{Q}  \\  
& $a_Q\,[10^{3}]$ & $b_Q\,[10^{3}]$ & $\theta_Q$ [\textdegree] & $x_{0,Q}-1226$ & $y_{0,Q}-1025$ &
<Res[$10^{-2}$\%]>$^{\star}$\\
\hline
$B$-band & 10.995$\pm$0.004 & 10.673$\pm$0.003 & -43.172$\pm$0.022 & -254.9$\pm$0.2 & 132.1$\pm$0.2 & 1.53$\pm$2.13\\
$V$-band & 9.688$\pm$0.002 & 10.129$\pm$0.003 & -43.894$\pm$0.018 & -283.9$\pm$0.2 & 191.7$\pm$0.2 & 1.55$\pm$2.51\\
$R$-band & 9.208$\pm$0.002 & 9.285$\pm$0.002 & -48.478$\pm$0.016 & -260.1$\pm$0.2 & 152.4$\pm$0.2 & 1.11$\pm$2.13\\
$I$-band & 8.556$\pm$0.002 & 8.643$\pm$0.002 & -49.274$\pm$0.014 & -297.9$\pm$0.2 & 177.3$\pm$0.2 & 0.91$\pm$2.46\\
%$H_{\alpha}$-band$^{\dagger}$ & 9.201$\pm$0.002 & 9.278$\pm$0.002 & -48.489$\pm$0.016 & -260.6$\pm$0.2 & 152.8$\pm$0.2 & -- \\
\hline
Filter & \multicolumn{6}{c}{U} \\
 & $a_U\,[10^{3}]$ & $b_U\,[10^{3}]$ & $\theta_U$ [\textdegree] & $x_{0,U}-1226$ & $y_{0,U}-1025$ &
<Res[$10^{-2}$\%]>$^{\star}$ \\%$^{\star}$
\hline
$B$-band & 10.045$\pm$0.003 & 10.528$\pm$0.004 & -0.361$\pm$0.011 & -226.4$\pm$0.2 & 214.7$\pm$0.2 & -0.18$\pm$2.75\\
$V$-band & 8.981$\pm$0.002 & 9.574$\pm$0.003 & -0.584$\pm$0.009 & -296.3$\pm$0.2 & 216.9$\pm$0.2  & 0.30$\pm$3.67\\
$R$-band & 8.781$\pm$0.002 & 9.071$\pm$0.003 & -2.644$\pm$0.008 & -278.9$\pm$0.1 & 195.5$\pm$0.1  & -0.02$\pm$3.43\\
$I$-band & 8.414$\pm$0.002 & 8.443$\pm$0.002 & -3.810$\pm$0.008 & -311.1$\pm$0.1 & 159.0$\pm$0.1 & 0.33$\pm$3.97\\

    \end{tabular}
    \begin{tablenotes}
      \small
     
      \item[$\star$] Median and median absolute deviation of the residual between original map and fitted paraboloid.
    \end{tablenotes}
\end{threeparttable}
}
\end{table*}

Although practical and convenient, analytic functions may not capture all small scale variations and might lead to artificial features that are not really present. We therefore also perform non-analytic fits in section~\ref{ap:inla} finding residuals that are 1-2 orders of magnitudes lower.

\subsection{Polarization map}\label{sec:pol}

\begin{figure*}
\includegraphics[trim={1.0cm 0 1.0cm 0 },clip,width=0.52\textwidth]{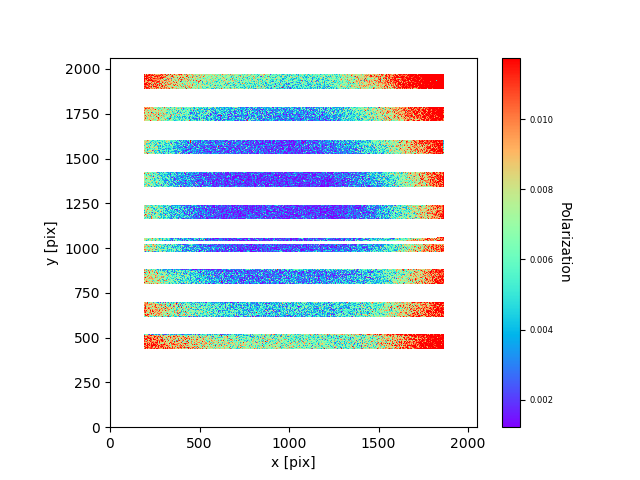}
\includegraphics[trim={1.0cm 0 1.0cm 0 },clip,width=0.52\textwidth]{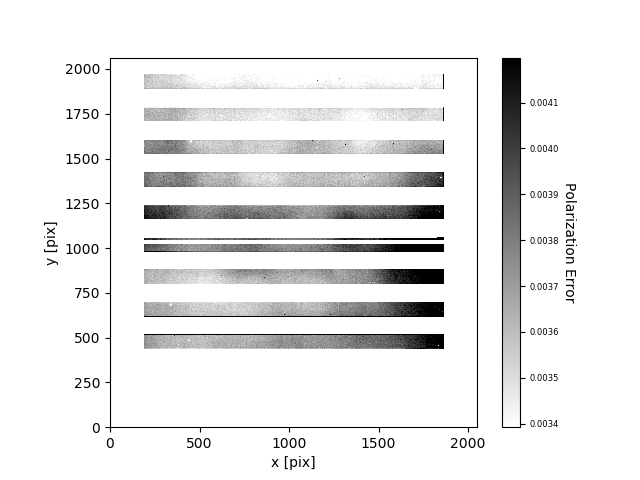}
\caption{\label{fig:polmap} Map of polarization degree (left) and error (right) for $V$-band calculated pixel by pixel for the moonlit sky with eight HWP angles, after correction for the background value.}
\end{figure*}

\begin{figure*}
\includegraphics[trim={1.0cm 0 1.0cm 0 },clip,width=0.52\textwidth]{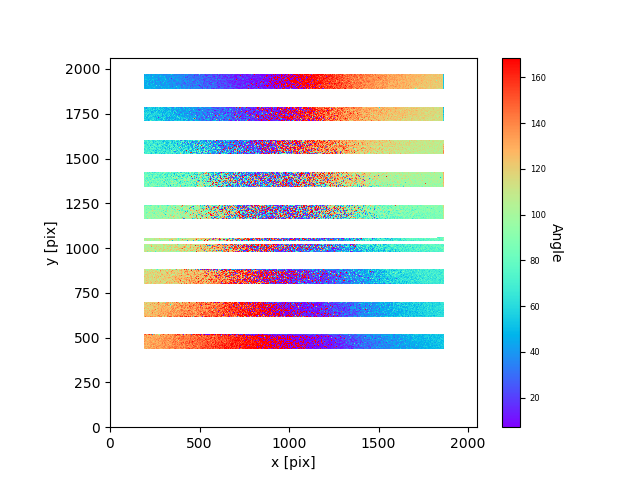}
\includegraphics[trim={1.0cm 0 1.0cm 0 },clip,width=0.52\textwidth]{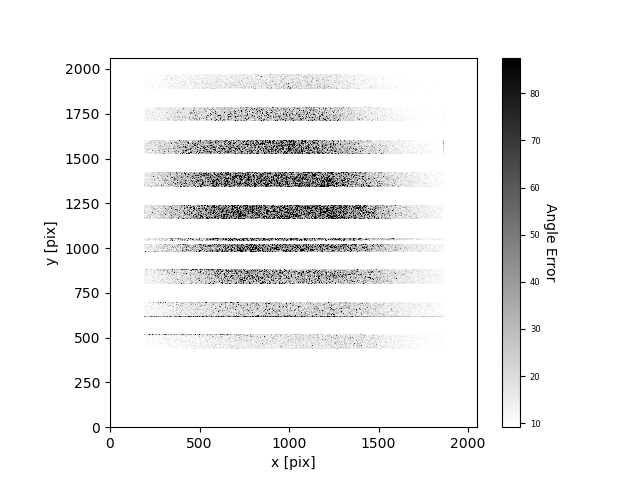}
\caption{\label{fig:angmap} Map of polarization angle (left) and error (right) for $V$-band calculated pixel by pixel for the moonlit sky with eight HWP angles, after correction for the background value.}
\end{figure*}

After correcting for the background polarization, we map the polarization and polarization angle along the FoV. We do this pixel by pixel and in of 30$\times$30 pixel bins. 
The results for this field instrumental polarization are shown in Figure~\ref{fig:polmap} and Figure~\ref{fig:angmap}. 

\begin{figure*}
\includegraphics[trim={1.0cm 0 1.0cm 0 },clip,width=1.0\textwidth]{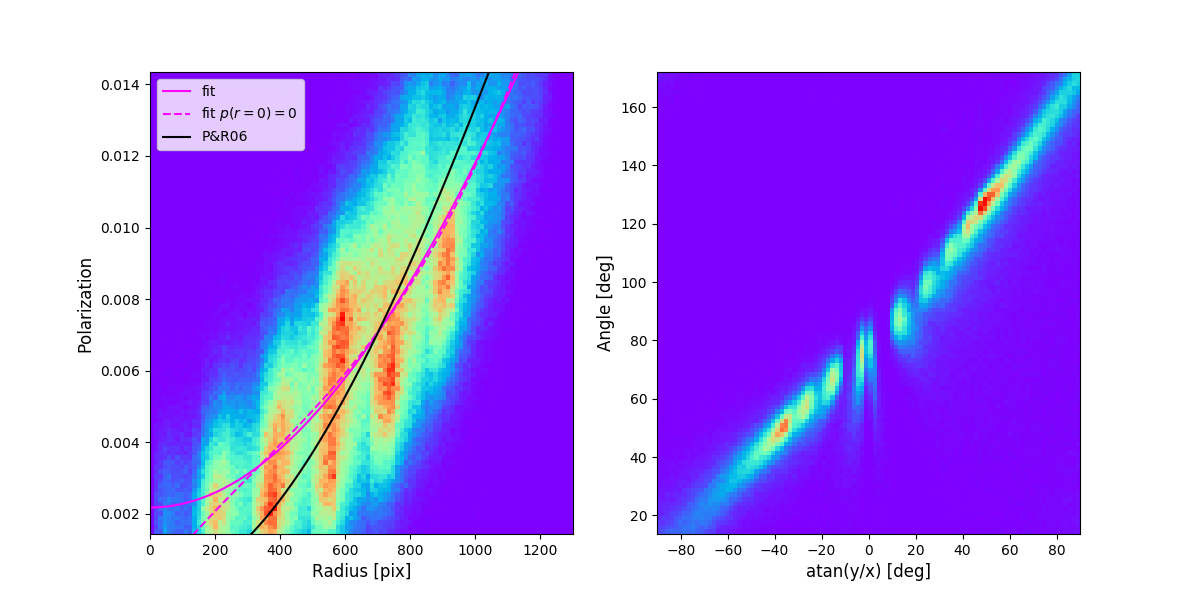}
\caption{\label{fig:polrad} 2D histogram of polarization vs radius (left) and polarization angle vs $\arctan{(y/x)}$ (right) for all pixels in the moonlit sky in $I$-band for FORS2. We show the radial fits of a cubic polynomial with/without forcing $P(r=0)=0$ as pink dashed/solid lines. The radial fit for FORS1 from PR06 is also shown in black.}
\end{figure*}

The radial pattern is quite clear in the polarization map increasing from nearly zero in the central telescope optical axis up to $\sim$1.4\% at the boundary depending on the filter. This can also be seen in Figure~\ref{fig:polrad} where we show the radial profile together with some polynomial fits. We find that the behaviour is well represented by a cubic polynomial:   
\begin{equation}\label{eq:radcubic}
     P(r) = m_1 r+m_2 r^2+m_3 r^3,
\end{equation}

\noindent with $m_{1,2,3}$ parameters for each filter given in Table~\ref{table:radcubic}, and with the radius, $r=\sqrt(x^2+y^2)$, given in pixels.
We note that these parameters are in good agreement with FORS1 (PR06), as expected if this large-scale pattern in the spurious polarization stems from the curved collimator lenses which are identical in FORS2. The distribution of polarization angle as a function of polar angle in the right Figure~\ref{fig:polrad} further emphasizes that the pattern is strongly radial.

\begin{table}[h!t]
\centering
 \caption{Parameters for cubic eq.~\ref{eq:radcubic} of polarization degree vs radius, $r=\sqrt(x^2+y^2)$, in pixels for $BVRI$ filters with respect to the optical axis ($x_0=1226,y_0=1025$pix)}.\label{table:radcubic}
 \begin{tabular}{c|ccc}
    \hline
\hline  
Filter & $m_1\,[10^{-5}]$ & $m_2\,[10^{-8}]$ & $m_3\,[10^{-12}]$ \\
\hline
$B$-band & 1.114$\pm$0.003 & -1.275$\pm$0.008 & 9.153$\pm$0.048 \\
$V$-band & 1.130$\pm$0.004 & -0.972$\pm$0.010 & 7.509$\pm$0.062 \\
$R$-band & 1.221$\pm$0.003 & -1.239$\pm$0.009 & 10.336$\pm$0.056 \\
$I$-band & 1.151$\pm$0.004 & -0.706$\pm$0.010 & 7.271$\pm$0.065 \\ 
    \end{tabular}
\end{table}

Although the radial 1D fit helps understanding the behaviour of the instrumental polarization as one moves towards the extremities of the FoV, a better approximation should take the azimuthal variation into account as well. This can be done directly with the corrected $Q$ and $U$ Stokes parameters calculated in the previous section, or, alternatively, we may perform a 2D fit to the polarization degree with a paraboloidal function that has the freedom of having the axes rotated, as follows:

\begin{eqnarray}\label{eq:radparab} 
     P(x,y) &=& \frac{[x'(x,y)]^2}{a_P^2} + \frac{[y'(x,y)]^2}{b_P^2} \\ \nonumber
     &=&\frac{[(x-x_{0,P})\cos\theta_P-(y-y_{0,P})\sin\theta_P]^2}{a_P^2} \\ \nonumber
     &+& \frac{[(x-x_{0,P})\sin\theta_P+(y-y_{0,P})\cos\theta_P]^2}{b_P^2}\nonumber
\end{eqnarray}
 
\begin{figure*}
\includegraphics[trim={1.4cm 1.5cm 1.2cm 1.5cm },clip,width=0.49\textwidth]{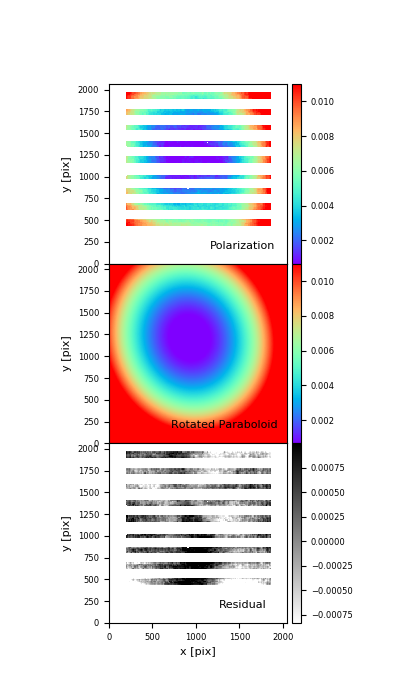}
\includegraphics[trim={1.4cm 1.5cm 1.2cm 1.5cm },clip,width=0.49\textwidth]{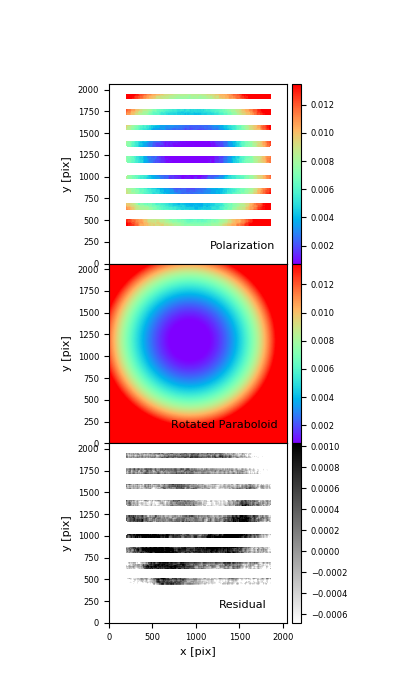}
\caption{\label{fig:polparab} Binned polarization map (top), rotated paraboloidal fit of eq.~\ref{eq:radparab} (middle) and residual between the two (bottom) for filters $V$ (left) and $I$ (right) of moonlit sky.}
\end{figure*}

The paraboloid is described by the free $a_P$ and $b_P$ parameters, whereas $x_{0,P}$ and $y_{0,P}$ represent the coordinates of the central $P=0$ position (with respect to the telescope optical axis) and $\theta_P$ determines the rotation of the paraboloid in the $xy$ plane. Such a fit and residual are shown in Figure~\ref{fig:polparab} for filters $V$ and $I$, while all final parameters are presented in Table~\ref{table:polparab}. We can see that the rotated paraboloid can remove most of the polarization pattern with residuals below $P\sim0.05$\%, which is within the estimated error (see appendix~\ref{ap:fourier}). The most asymmetrical pattern is found for $B$ and $V$ bands in agreement with the results of PR06 for FORS1. These asymmetries can be seen directly in the polarization map obtained from the corrected $Q$ and $U$ or via non-analytic fits, as shown in next subsection. The origin of this asymmetrical polarization is unknown.  

\begin{table*}[h!t]
\centering
\begin{threeparttable}
 \caption{Parameters for the rotated paraboloid of eq.~\ref{eq:radparab} for field polarization for $BVRI$ filters. The origin of the$x,y$  coordinate system is taken at the bottom left of the combined image}.\label{table:polparab}
 \begin{tabular}{c|cccccc}
    \hline
\hline  
Filter & $a_P\,[10^{3}]$ & $b_P\,[10^{3}]$ & $\theta_P$ [\textdegree] & $x_{0,P}-1226$ & $y_{0,P}-1025$ & <Res[$10^{-2}$\%]>$^{\star}$\\
\hline
$B$-band & 10.121$\pm$0.003 & 10.510$\pm$0.004 & 10.027$\pm$0.344 & -52.34$\pm$0.19 & 147.02$\pm$0.19  & 0.36$\pm$3.47\\
$V$-band & 8.493$\pm$0.002 & 10.634$\pm$0.004 & 20.569$\pm$0.057 & -87.20$\pm$0.16 & 178.09$\pm$0.20 & 0.91$\pm$3.67\\
$R$-band & 8.841$\pm$0.002 & 9.092$\pm$0.003 & -20.913$\pm$0.401 & -72.18$\pm$0.15 & 148.77$\pm$0.15 & 0.61$\pm$3.42\\
$I$-band & 8.181$\pm$0.002 & 8.584$\pm$0.002 & -69.385$\pm$0.229 & -94.56$\pm$0.15 & 150.05$\pm$0.13 & 0.85$\pm$3.36\\ 
%$H_{\alpha}$-band$^{\dagger}$ & 8.833$\pm$0.002 & 9.086$\pm$0.003 & -21.486$\pm$0.400 & -72.44$\pm$0.15 & 148.79$\pm$0.15 & -- & \\
    \end{tabular}
     \begin{tablenotes}
      \small
      \item[$\star$] Median and median absolute deviation of the residual between original map and fitted paraboloid. The median residuals correspond to 0.061$\sigma_B$, 0.056$\sigma_V$, 0.172$\sigma_R$ and 0.381$\sigma_I$, where $\sigma$ is the characteristic median error in each filter (see Table~\ref{table:error}). 
    \end{tablenotes}
\end{threeparttable}
\end{table*}

\subsection{Non-analytic map: INLA model}\label{ap:inla}

We use here a non-analytic fitting procedure to obtain maps of Stokes parameters, $Q$ and $U$, the polarization degree, $P$, and angle, $\chi$, of the FORS2 instrument optics in $BVRI$ obtained from the same moonlit blank fields investigated in the previous sections. In principle, this allows for a reconstruction that is less subject to forced features from the assumed analytic function. We use a method based on Gaussian Markov Random Fields, to model the underlying continuous spatial field with hyper-parameters fitted with a Bayesian inference algorithm known as Integrated Nested Laplace Approximation \citep[INLA,][]{Rue17}. The method offers the additional advantage of full posterior distributions to properly model uncertainties. INLA has been successfully applied in many different areas including extended sources in astronomy \citep{Gonzalez19}.  

\begin{figure*}
\includegraphics[trim={1.4cm 1.5cm 1.2cm 1.5cm },clip,width=0.49\textwidth]{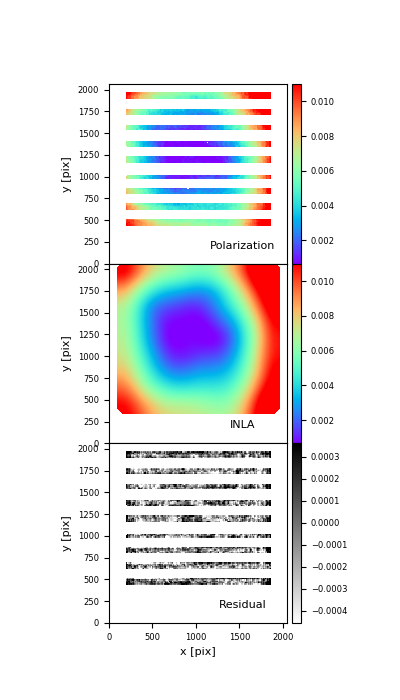}
\includegraphics[trim={1.4cm 1.5cm 1.2cm 1.5cm },clip,width=0.49\textwidth]{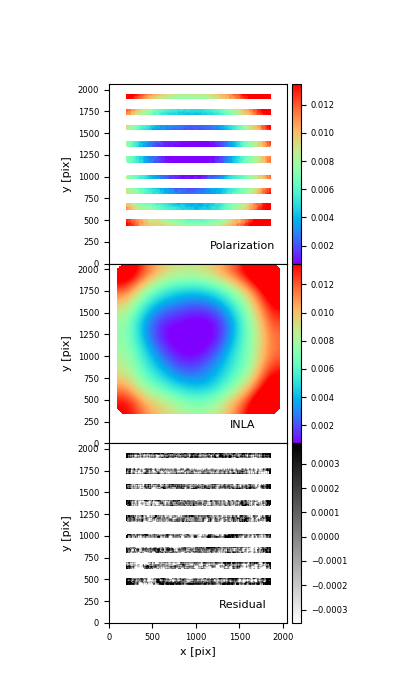}
\caption{\label{fig:polinla} Binned polarization map (top), INLA fit (middle) and residual between the two (bottom) for filters $V$ (left) and $I$ (right) of moonlit sky.}
\end{figure*}

Although at first the INLA reconstructions of $Q$ and $U$ look quite similar to the analytic hyperbolic paraboloid, the residuals in $B$-band are at least one order of magnitude smaller and show no clear patterns, as opposed to the analytic fits. The advantage of the INLA fits becomes more evident in the polarization fits shown in Figure~\ref{fig:polinla}, where the asymmetric behaviour is clearly visible in the reconstruction and it is confirmed in the rather flat residual map. We present the median and deviation of all residual maps in Table~\ref{table:resinla}. We also release all final INLA reconstructions\footnote{See   \href{https://github.com/gongsale/FORS2-INSTPOL/}{https://github.com/gongsale/FORS2-INSTPOL/}}.

We have presented in this section analytic and non-analytic spatial corrections for the spurious instrumental linear polarization of broad-band filters. In the following sections we will validate these results with other datasets.

\begin{table*}[h!t]
\centering
% \scalebox{0.7}{
\begin{threeparttable}
 \caption{Median and median absolute deviation of the residual maps for the INLA fits of $Q$, $U$, $P$ and $\chi$.}\label{table:resinla}
 \begin{tabular}{c|cccc}
    \hline
\hline  
Filter & $B$-band & $V$-band & $R$-band & $I$-band\\
\hline
<Res(Q) [$10^{-2}$\%]> & 0.027$\pm$1.500 & 0.159$\pm$1.460 & 0.037$\pm$1.419 & 0.017$\pm$1.337 \\
<Res(U) [$10^{-2}$\%]> & -0.096$\pm$1.454 & -0.147$\pm$1.626 & -0.032$\pm$1.528 & -0.056$\pm$1.363 \\
<Res(P) [$10^{-2}$\%]>$^{\Diamond}$ & 0.018$\pm$1.514 & -0.654$\pm$1.652 & -0.003$\pm$1.531 & 0.026$\pm$1.396 \\
<Res($\chi$) [$10^{-1}$ deg]> & 0.116$\pm$8.455 & 0.337$\pm$6.650 & 0.240$\pm$4.776 & 0.284$\pm$5.678 \\
    \end{tabular}
    \begin{tablenotes}
      \small
      \item[$\Diamond$] These median polarization degree residuals correspond to 0.003$\sigma_B$, -0.040$\sigma_V$, -0.001$\sigma_R$ and 0.012$\sigma_I$ for $BVRI$ respectively, where $\sigma$ is the median error of each filter from Table~\ref{table:error}.
    \end{tablenotes}
\end{threeparttable}
%}
\end{table*}

%% file: sec_5_M30.tex
In this section, we use independent observations of the stellar cluster M30 to test the maps of instrumental polarization found in the previous section. M30 was observed with several ditherings in broad-band filters $RI$ and in the narrow $H_{\alpha}$ filter. Containing hundreds of stars across the field, we can validate the spatial instrumental pattern by performing stellar photometry in the images taken in each HWP angle and then applying the same formalism of section~\ref{sec:theory} to measure polarization of each individual star. This independent method offers some benefits: while we can test if the spurious instrumental polarization persists in a field with mostly unpolarizaed sources and no strong background polarization, it also permits to study the effects of ghosting, vignetting, Point Spread Function (PSF) variations across the field and between o- and e-beams \citep[see e.g.][]{Clemens12}. The disadvantages, besides the difficulties of performing crowded stellar photometry, is the possible presence of varying interstellar dust polarization across the field, as well as foreground and background stars with polarization degrees that differ from the M30 member stars.

\subsection{Field polarimetry}
\begin{figure*}%[t!]
  		\includegraphics[trim={0.6cm 0.0cm 0.6cm 0cm },clip,width=0.49\linewidth]{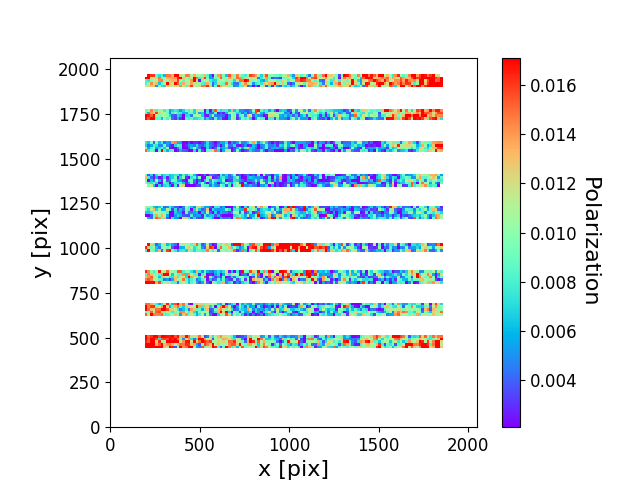}
  		\includegraphics[trim={0.6cm 0.0cm 0.6cm 0cm },clip,width=0.49\linewidth]{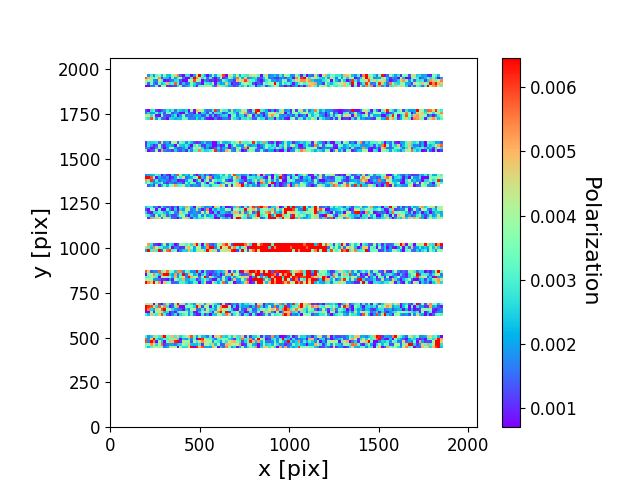}
  		\caption{Polarization map for a field with M30 in $R$ for binned (30$\times$30 pix) boxes (left) and after spatial field correction in $Q$ and $U$ using eq.~\ref{eq:hypparab} (right). Note the residual central polarization from the cluster center.}   
  		\label{fig:M30-Rbin}
%	\end{minipage}
\end{figure*}

As a first simple step, we use the same methodology as for the moonlit sky (section~\ref{sec:instpol_moon}) applied now to the M30 cluster: we bin each image in boxes of 30$\times$30 pixels and calculate $Q$, $U$, $P$ and $\chi$. It is important to note that in this first step we do not perform any stellar photometry. Even though the S/N at each pixel is not as high as for the bright moonlit sky, Figure~\ref{fig:M30-Rbin} shows a similar radial instrumental pattern in $R$-band (after correction for the median background $Q$ and $U$). This pattern is confirmed for all pointings of $R$ and $I$. If we apply the spatial correction for the $Q$ and $U$ maps that we obtained from the hyperbolic paraboloid fits (see Table~\ref{table:hypparab}), we find a roughly flat map, except for the center where the crowded nucleus of the stellar cluster leaves a residual polarization. This demonstrates the validity of our instrumental correction.

\subsection{Stellar polarimetry}

Now we turn to the more difficult task of doing photometry for crowded fields of each o- and e-beam images of all HWP angles at different offsets. Due to the high density of sources, aperture photometry for stellar clusters is generally inaccurate and PSF modeling is required. For this we use a general PSF template across the field built from a set of bright stars. To avoid systematic biases, we use the same PSF for all HWP angle exposures of a given offset and filter. The sources are found by selecting significant brightness peaks above the background, which are then fitted to the PSF model and subtracted. The process is repeated to find any left-over sources. All steps are included in the packages {\sc photutils}\footnote{\url{https://doi.org/10.5281/zenodo.2533376}} in {\sc PYTHON} or {\sc starfinder} in {\sc IDL} \citep{Diolaiti00}. It is worth mentioning that, when doing photometry, there is a concern that the background subtraction may eliminate some of the polarization pattern. Our tests with/without subtraction are consistent throughout. 

\begin{figure*}%[t!]
  		\includegraphics[trim={0.2cm 0.0cm 0.6cm 0cm },clip,width=0.33\linewidth]{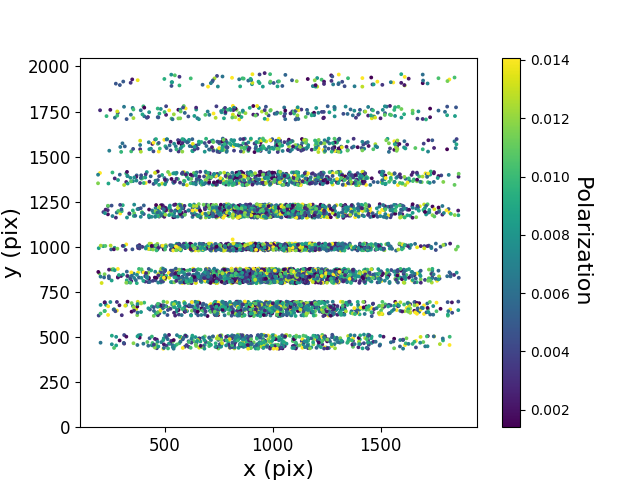}
  		\includegraphics[trim={0.6cm 0.0cm 0.6cm 0cm },clip,width=0.33\linewidth]{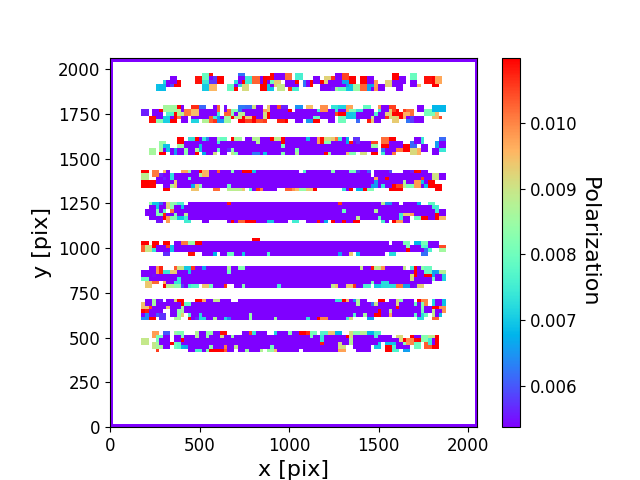}
  		\includegraphics[trim={0.6cm 0.0cm 0.6cm 0cm },clip,width=0.33\linewidth]{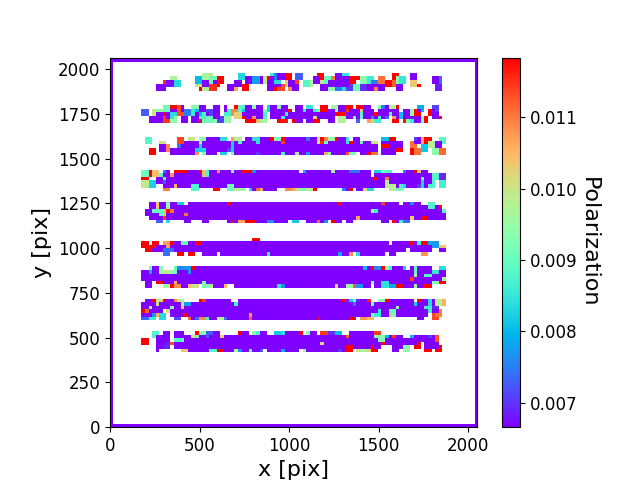}
  		\caption{Individual stellar polarization map for M30 of three combined pointings in $R$ (left) and binned in 20$\times$20 boxes (middle) and after spatial field correction in Q and U using eq.~\ref{eq:hypparab} (right). The binned maps have characteristic polarization error of $\sim0.4\%$.}   
  		\label{fig:M30-Rstar}
%	\end{minipage}
\end{figure*}

To match the star catalogs of different HWP angles, we look for the closest star within 1 pixel in both $x$ and $y$ positions. We then calculate the $Q$ and $U$ Stokes parameters from eq.~\ref{eq:QU} and the polarization from eq.~\ref{eq:pol}) for each star. We show in the left Figure~\ref{fig:M30-Rstar} the resulting star polarization degree for M30 in $R$-band. This figure includes three different pointings with the center slightly offset and a total of more than 13000 star positions. Because of the multiple pointings of the same field, many stars are repeated, yet at various locations in the field. The scatter in the polarization of stars is very large; we therefore bin the individual star $Q,U$ values into boxes of 20$\times$20 pixel boxes rejecting 2$\sigma$ outliers and calculate the polarization within each box, as shown in the middle plot. We can see here a radial pattern similar to Figure~\ref{fig:M30-Rbin}, which we correct with the hyperbolic paraboloid function and parameters found with the moonlit observations, removing thus some of the spurious edge polarization and obtaining a flatter pattern (right plot). This is also consistent with the moonlit sky approach, and argues for an instrumental field polarization that is valid for both, polarized and unpolarized, incoming light.

We emphasize however that the correction is not optimal with a large scatter in the distribution of degrees of star polarization. This may come from the large photometric uncertainties obtained from PSF photometry in very dense clusters. Doing aperture photometry instead, or restricting the analysis to only a fraction of stars with high SNR, does not reduce the scatter. This was
already noted by \citet{Clemens12}, where they ultimately perform precision photometry by first removing stars from around a target star using
PSF stellar fits, then doing aperture photometry on the target star, and iterating this procedure for new target stars. This approach is outside the scope of this paper.

\subsection{The case of the narrow $H_{\alpha}$ filter}

\begin{figure}%[t!]
  		\includegraphics[trim={0.6cm 0.0cm 0.6cm 0cm },clip,width=0.49\linewidth]{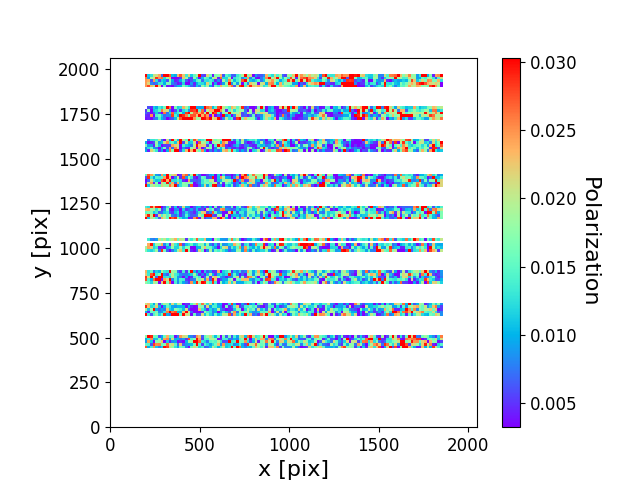}
  		\includegraphics[trim={0.6cm 0.0cm 0.6cm 0cm },clip,width=0.49\linewidth]{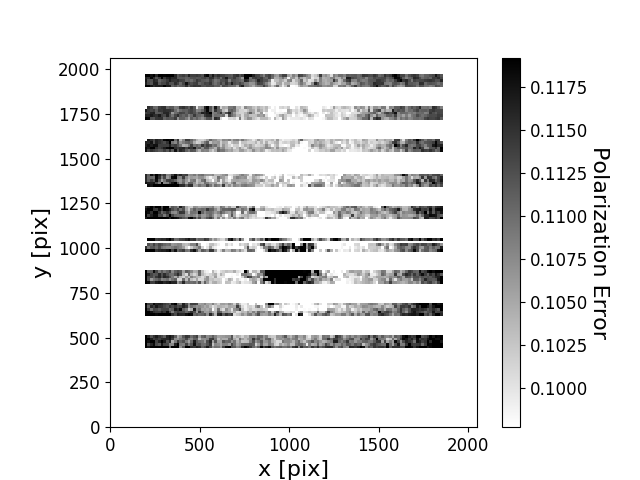}
  		\caption{Binned $H_{\alpha}$ polarization map for M30 at a single pointing (left) and error (right).}   
  		\label{fig:M30-Hbin}
%	\end{minipage}
\end{figure}

At first sight, one may expect that the behaviour of the spatial instrumental pattern extends to other wavelengths not explored with the bright sky observations, so that a simple interpolation of the paraboloid parameters to the effective wavelengths of other filters should suffice. However, the narrow-band interference filters are in the converging beam, instead of the collimator, as is the case for the broad-band filters. If the primary spurious pattern is due to the curved lenses in the collimator, one may wonder if this affects the instrumental pattern. We try here to explore this with our observations of M30 in the $H_{\alpha}$ filter at four HWP angles and three different pointings. By doing a similar analysis as with $R$ and $I$, we first find that in a binned analysis (see Figure~\ref{fig:M30-Hbin}), there is no apparent radial pattern. However, this is within the inferred error which is of the order of $\sim$10\% in polarization. Although quite low, this SNR is only roughly a factor of 2 lower than for the broad-band filters, where the pattern is clearly seen. This is confirmed for all three M30 pointings and for an additional standard star field\footnote{Vela1 at Modified Julian Date = 58281, see \S~\ref{sec:discussion} for the analysis of standard star fields}, albeit all with low SNR. 

\begin{figure}%[t!]
  		\includegraphics[trim={0.6cm 0.0cm 0.6cm 0cm },clip,width=0.49\linewidth]{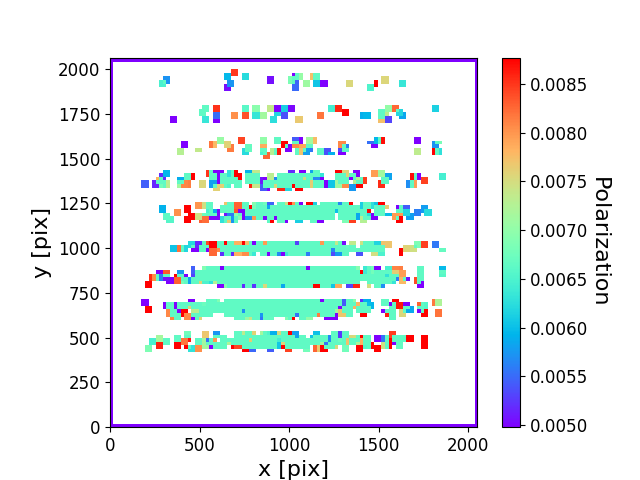}
  		\includegraphics[trim={0.6cm 0.0cm 0.6cm 0cm },clip,width=0.49\linewidth]{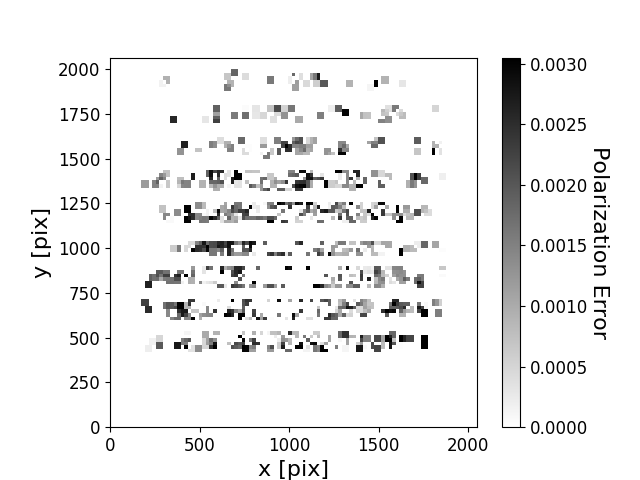}
  		\caption{$H_{\alpha}$ star polarization degree map for M30 in three pointings binned in 20$\times$20 boxes (left) and respective error (right)}   
  		\label{fig:M30-Hstar}
%	\end{minipage}
\end{figure}

Using stellar PSF photometry, our results look similar to the broad-band filters (see Figure~\ref{fig:M30-Hstar}): the polarization degree seems to increase towards the outer parts of the CCD in accordance to the instrumental pattern found at other wavelengths of broad-band filters. However, the degree of polarization is lower compared to the broad-bands (up to $\sim 0.9\%$ versus $1.4\%$), and when correcting for the instrumental pattern found previously, we do not obtain lower residuals in the polarization map. Although the SNR is low, even after binning individual $Q,U$ star values, the scatter is large suggesting again that a more robust photometry technique should be applied, or a higher SNR is needed.

%% file: sec_6_discussion.tex
In this section, we briefly discuss some biases that may affect the stability of the instrumental polarization pattern found in the previous sections. 

Firstly, there is a known effect of cross-talk in FORS2 in which linear polarization creates a non-negligible circular polarization \citep{Bagnulo09}. On-axis, the induced circular polarization is quite small for unpolarized sources ($<<0.01$\%) growing for highly polarized objects ($\sim0.5$\% for 10\% polarization). Since the spatial linear instrumental polarization increases with off-axis distance, it is expected that this cross-talk might increase as well. In fact, these authors (see their Table~1) find an increase in the circular polarization cross-talk by a factor of roughly $\sim$1.2 to the edge of the detector. This corresponds surprisingly well with the radial increase in linear polarization that we find in this work, which could thus partly explain their findings. Although we have neglected the study of circular polarization here, the question arises if the found instrumental pattern changes, for instance, with varying background linear polarization. Although the study of unpolarized cluster stars in a moonless night (see previous section) suggests that the spatial instrumental pattern is global, more observations with different background polarizations are advisable.  

\begin{figure*}%[t!]
\centering
  		\includegraphics[trim={0.4cm 0.2cm 0.4cm 0cm },clip,width=0.85\linewidth]{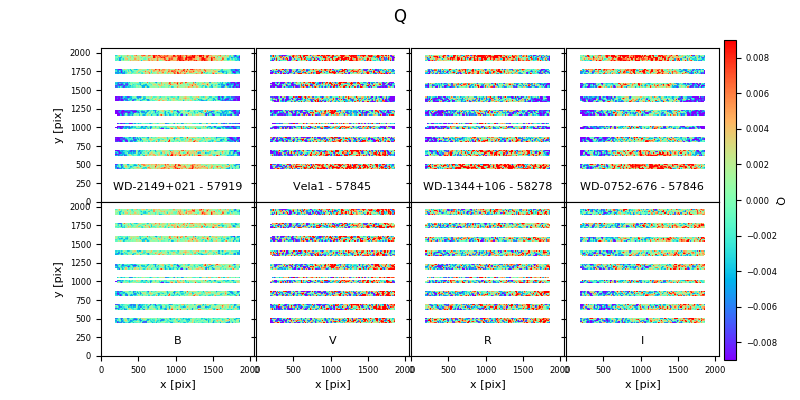}
  		\includegraphics[trim={0.4cm 0.2cm 0.4cm 0cm },clip,width=0.85\linewidth]{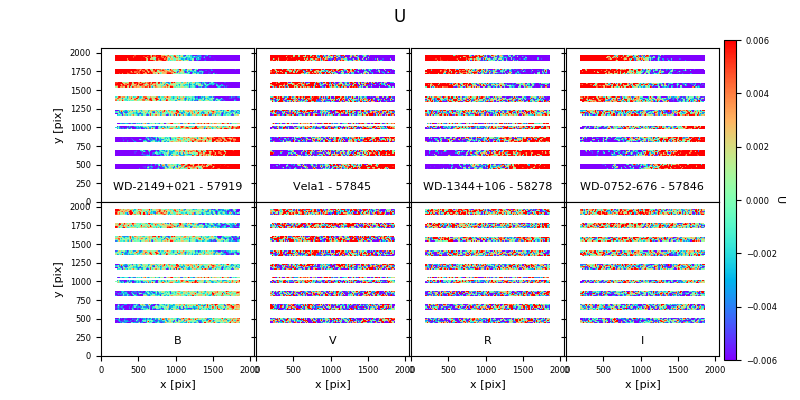}
  		\includegraphics[trim={0.4cm 0.2cm 0.4cm 0cm },clip,width=0.85\linewidth]{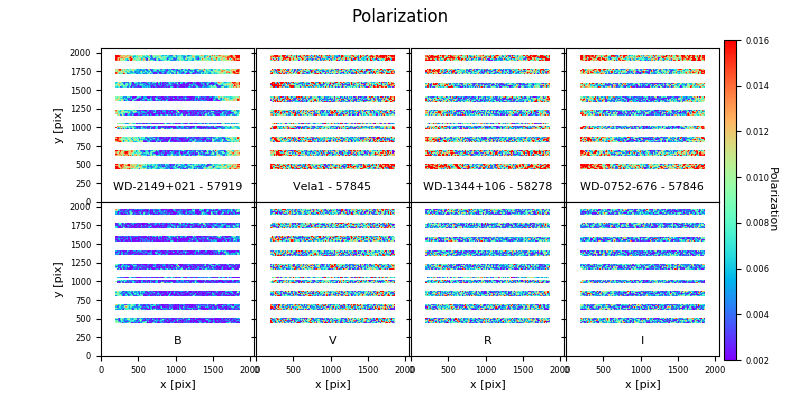}
  		\caption{Binned $Q$ (upper two rows), $U$ (middle two rows) and $P$ maps (bottom two rows) after correction for median $Q_0,U_0$ values (upper) and after correction of instrumental field polarization from paraboloid fits of \S~\ref{table:hypparab} (lower) for standard star fields of a) WD-2149+021 at Modified Julian Date (MJD) of 57919 in $B$-band, b) Vela1 at MJD=57845 in $V$-band, c) WD-1344+106 at MJD=58278 in $R$-band and d) WD-0752-676 at MJD=57846 in $I$-band.}   
  		\label{fig:comp}
  		\end{figure*}

Furthermore, there may be some concerns about telescope flexure: it is known that there is image motion due to instrument flexure under gravity below 0.25 pixel over a one hour exposure with the SR collimator for zenith distances less then 60\textdegree (see FORS2 manual). Although this is rather small, the effects on the spurious linear polarization pattern are still unknown. 

Finally, a natural concern is that our corrections may evolve with time. For instance, as coatings age or get changed, as mirrors get re-coated, or as the instrument is unmounted and mounted again, there may be variations in the induced instrumental pattern. \citet{Wiersema18} find for example a time dependent effect on the calibration of the instrument EFOSC2 due to oxidation, dust, and perhaps other factors, settling on the tertiary mirror of the NTT. 

To gauge some of these concerns, since our moonlit sky observations were taken only at one pointing, we investigate here the instrumental polarization for other targets with different Moon separations and therefore varying background polarization degrees and angles. We study several fields observed with FORS2-IPOL in $BVRI$ available from the ESO archive. The data were obtained at different dates and times and thus at varying background linear polarizations, telescope elevations, among others. Most of these data come from standard star fields observed in linear polarimetry mode and have much less signal-to-noise compared to our moonlit sky observations. In Figure~\ref{fig:comp}, we show binned $Q$, $U$ and $P$ for four different fields at different epochs. Albeit with lower SNR, we can recognize the pattern found previously. When we correct with the analytic paraboloid functions, we obtain residual maps of the order of $\sim10^{-4}$, lower than the typical intrinsic error of $\sim10^{-2}$.

\begin{figure*}%[t!]
  		\includegraphics[trim={0cm 0.6cm 0.9cm 0cm },clip,width=0.33\linewidth]{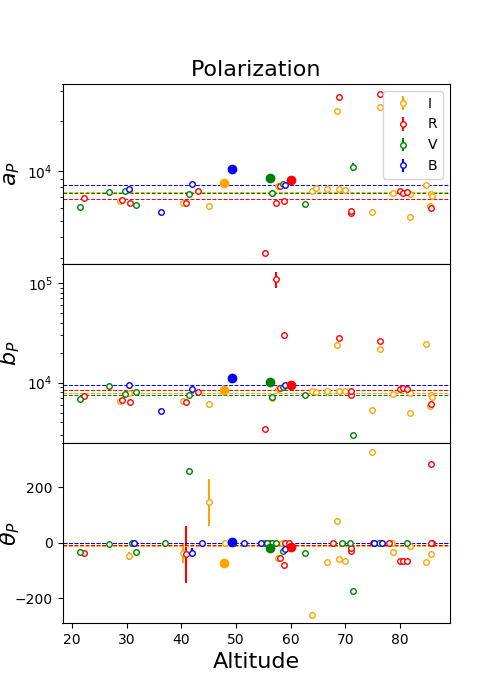}
  		\includegraphics[trim={0cm 0.6cm 0.9cm 0cm },clip,width=0.33\linewidth]{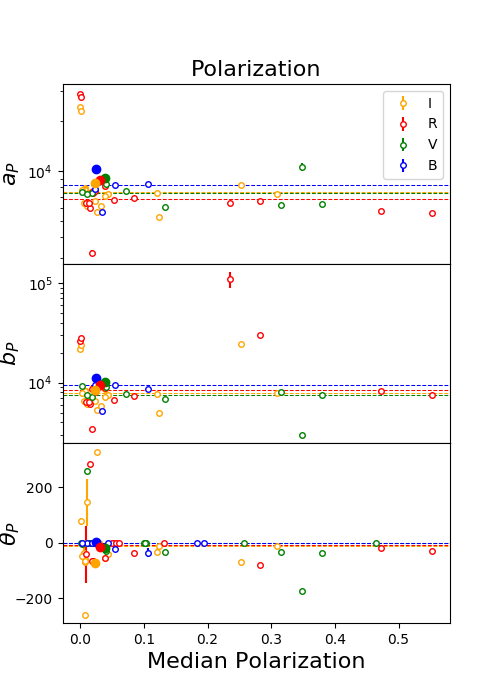}
  		\includegraphics[trim={0cm 0.6cm 0.9cm 0cm },clip,width=0.33\linewidth]{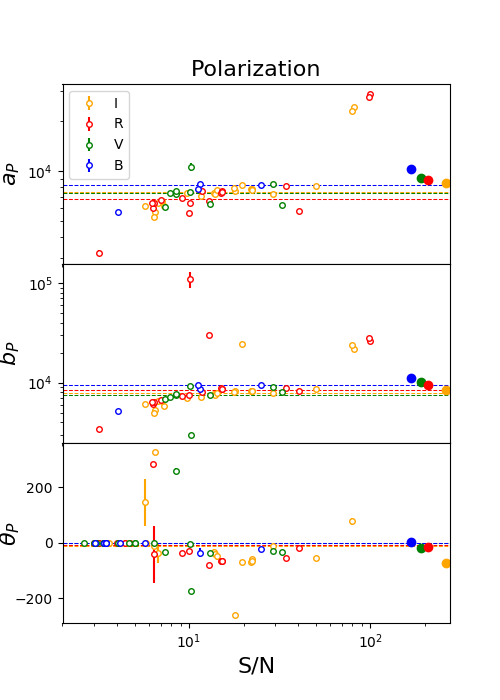}
  		\caption{Paraboloid fit parameters $a_P$ (top), $b_P$ (middle) and $\theta_P$ (bottom) of eq.~\ref{eq:radparab} for the polarization degree of multiple FORS2 fields as a function of target altitude (left), median background polarization degree (middle) and intensity SNR (right). Different filters are shown in colors and filled circles represent our moonlit sky observations. Dashed horizontal lines are the median of the parameters per filter.}   
  		\label{fig:evol}
  		\end{figure*}

Furthermore, if we perform individual paraboloid fits for each of the cases we retrieve from the archive, we obtain quite consistent fit parameters as shown in Figure~\ref{fig:evol}, finding no evidence for variation with observation date nor altitude/azimuth of the target. The data we investigate spans observations from 2011 until 2018 suggesting that there is not a significant time evolution of the instrumental linear polarization. This is confirmed by the similar pattern found already by PR06 for FORS1 (see \S~\ref{sec:instpol_moon}). We also do not find changes with background sky polarization (which is always corrected for prior to the fit), but there is some scatter and only few observations with high polarization. We also see that when the intensity has SNR$\lesssim10$, the obtained parameters drift off from the median values for all $Q$, $U$ and $P$ parameters. This finding suggests a minimum signal-to-noise cutoff for reliable spatial linear polarization studies. For low polarization degrees, i.e. $P\lesssim0.03$, this SNR corresponds to $P/\sigma_P\lesssim0.5$, a value much smaller than the typical $3\sigma$ limit to consider real linear polarization detections \citep[e.g.][]{Manjavacas17}.    Finally, it is important to mention that although the fit parameters are mostly consistent, a couple of outliers exist, even though they have decent SNR and low background polarization. The origin of this discrepancy is unknown. 

Therefore, the changes in the spatial instrumental linear polarization need to be further studied with high signal-to-noise observations, ideally with moonlit sky observations (as in \S~\ref{sec:instpol_moon}) at different telescope elevation angles and Moon separations, i.e. at different background sky polarizations. This would better clarify the stability of the instrumental polarization pattern under different observing conditions. By adding circular polarization measurements, it will be possible to address the question of possible cross-talk in a more systematic way.

%% file: apB_Rayleigh_moon.tex
The method used in this work to calculate the instrumental polarization relies on the assumption that it is null in the optical axis and that we can therefore correct it, and that the night sky polarization is constant within the field of FORS2. To study this further, we use here a simple model of single Rayleigh scattering from the Moon \citep[e.g.][]{strutt,harrington11}, simply described as:

\begin{equation}
P = \frac{\sin^2(\gamma)}{1+\cos^2(\gamma)},
    \end{equation}
    where $\gamma$ is the angle between the Moon and the target pointed by the telescope. 

\begin{figure}
\centering
\includegraphics[trim={0.5cm 0 1.cm 0 },clip,width=0.5\textwidth]{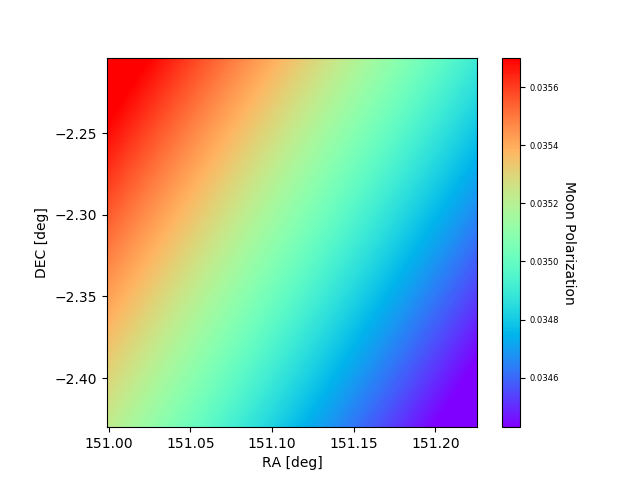}
\caption{\label{fig:moon} Simulated monochromatic single Rayleigh scattering model of the Moon at the location and time of our blank target observation with the FoV of FORS2. Polarization degree here is ($3.51\pm0.05$)\%.}
\end{figure}

Although our observations were taken on a full Moon night, throughout the duration of all exposures in all filters, the angle from the Moon changed between 17.69\textdegree and 18.31\textdegree  which corresponds to a raw single Rayleigh scattering polarization of $\sim3.51-4.69$\%.% 
This is somewhat higher than the values we measure and quote in section~\ref{sec:centralpol}. We assume that this comes from our simplistic mono-chromatic model that does not take into account multiple scattering and proper atmospheric models. 
We show one of our simulations in Figure~\ref{fig:moon}. It is important to see that in this model, the polarization degree from Moon scattering does not change more than 0.06\% throughout the field of FORS2, a value that is below the expected polarization we can measure (see appendix~\ref{ap:fourier}). Even if the real pattern is more complicated when one goes beyond simple Rayleigh scattering \citep[e.g.][]{Gal01,Berry04}, we do not expect the polarization to change substantially within the small angular scales of the FoV. Findings of constant sky polarization in small fields have been observed by others \citep{Wolstencroft73}.

%% file: ap_Fourier.tex
The Fourier decomposition allows for the identification of different sources of error, including instrumental components. The relation between the normalized flux differences and the Stokes parameters, defined in 
eq.(\ref{eq:FNi}) allows for such error analysis based on the Fourier transform.  
Indeed, in the case of observations using N=4, 8, 12 and 16 position angles of the HWP with a constant interval of $\pi/8$, the normalized flux differences,  $F_i$ in eq.(\ref{eq:FNi}),  can be rewritten as

\begin{equation}\label{eq:FS}
F_i = Q_{\rm o} +  \; 
         \sum_{k=1}^{N/2} Q_k \cos{\left(k\frac{2\pi\,i}{N}\right)} + 
          U_k \sin{\left(k\frac{2\pi\,i}{N}  \right) } \, , 
\end{equation}

\noindent where $Q_{0}, \, Q_k,\,$ and $U_k$ are the Fourier coefficients.   

\noindent From the Fourier coefficients, we can obtain the polarization of each harmonic, i.e. the polarization spectrum:

\begin{equation}
P_k = \sqrt{Q_k^2+U_k^2}.    
\end{equation}

\noindent Comparing the above formulae with equations~\ref{eq:QU} and \ref{eq:FNi}, it is possible to understand that the linear polarization signal is given by the $k=N/4$ harmonic, whereas the other components are related to different effects like instrumental imperfections and noise \citep{1996Fendt}. 

%%% Figure with Fourier coefficients 
\begin{figure*}
\centering
\includegraphics[trim={2.5cm 2cm 2.5cm 0},
clip,width=1.0\textwidth]
{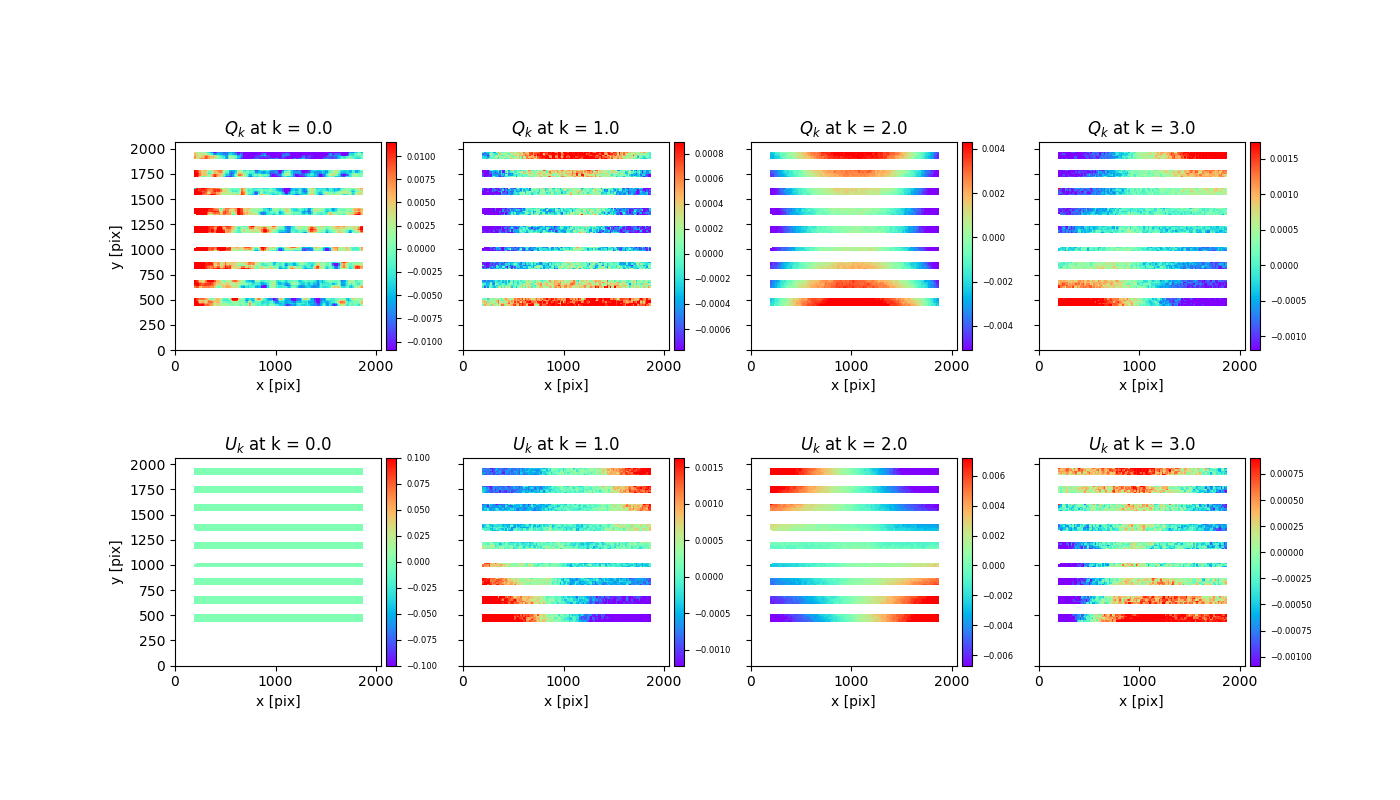}  
\caption{\label{fig:fc_b} Binned (30$\times$30 pixel) Fourier $Q_k$ and $U_k$ coefficients for a moonlit blank sky field in $B$-band. Note that the color scale in each plot is different. All values have been corrected for median background $Q_{k,B}, U_{k,B}$ values.}
\end{figure*}

\begin{figure*}
    \centering
\includegraphics[trim={2.5cm 2cm 2.5cm 0 },
clip,width=1.0\textwidth]
{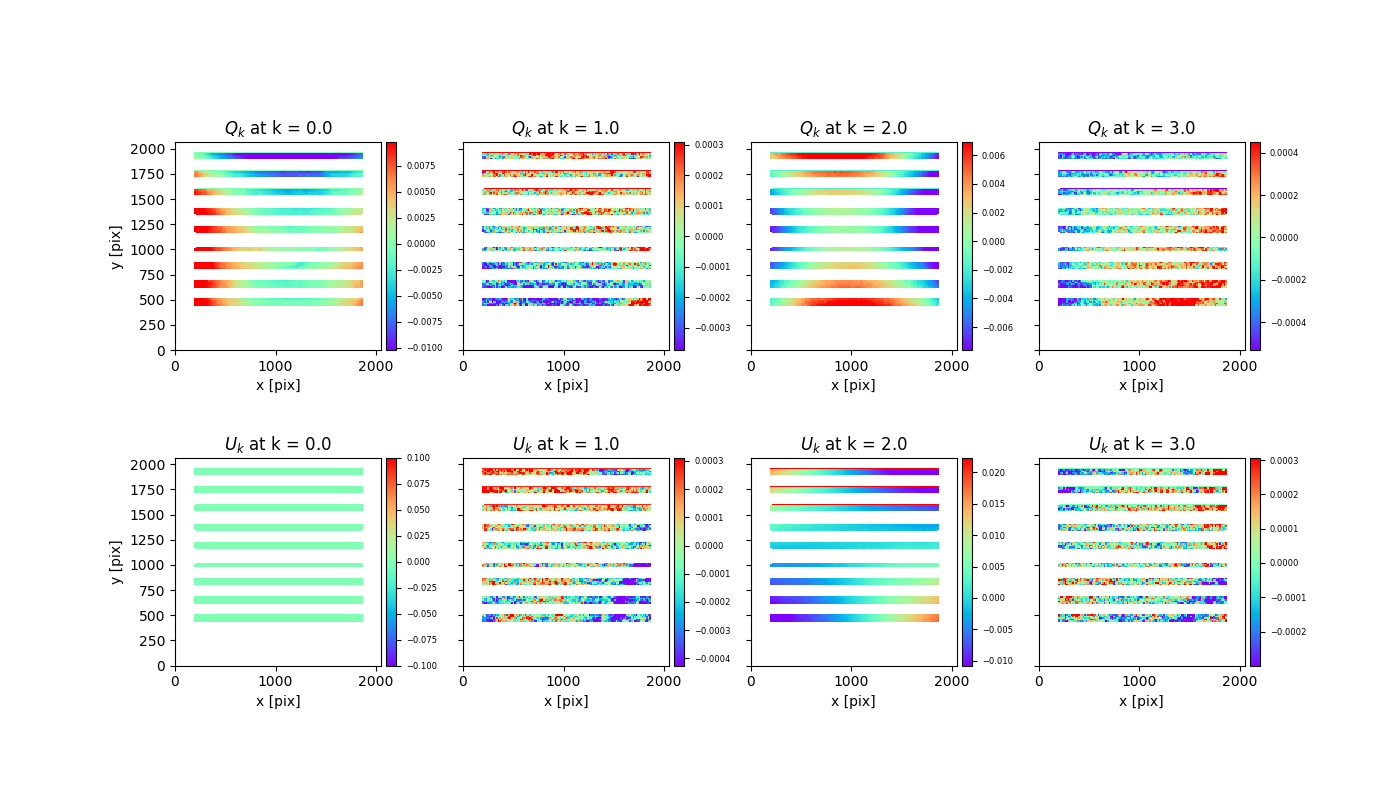}
\caption{\label{fig:fc_I} Binned (30$\times$30 pixel) Fourier $Q_k$ and $U_k$ coefficients for a moonlit blank sky field in $I$-band. Note that the color scale in each plot is different. All values have been corrected for median background $Q_{k,B}, U_{k,B}$ values.}
\end{figure*}

\begin{figure*}[h!t]
\centering
\includegraphics[trim={4.5cm 1.5cm 3.5cm 1.5cm},
clip,width=0.4\textwidth]
{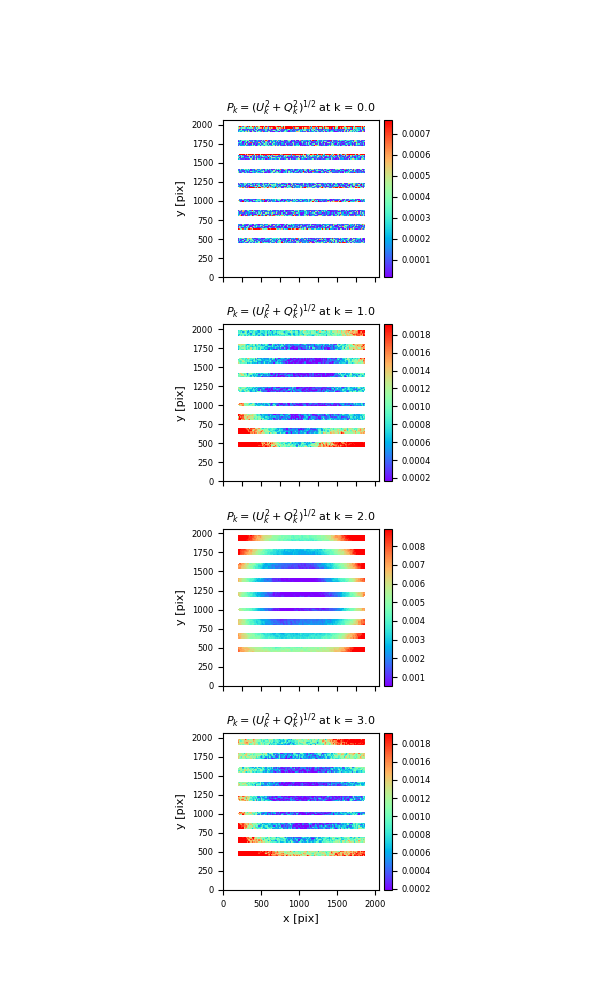} 
\includegraphics[trim={4.5cm 1.5cm 3.5cm 1.5cm},
clip,width=0.4\textwidth]
{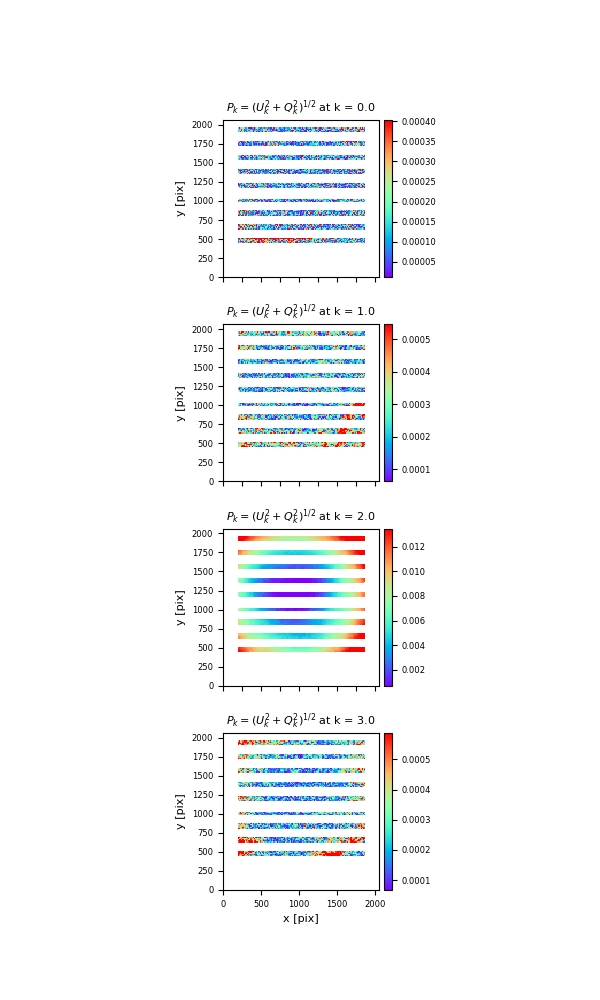}
\caption{\label{fig:p_spec} Binned (30$\times$30 pixel) Fourier power spectra $P_k = \sqrt{Q_k^2+U_k^2}$ for a moonlit blank sky field in $B$-band (left) and $I$-band (right). This is \emph{after} flat correction and median background $Q_{k,B},U_{k,B}$ correction.}
\end{figure*}

\noindent The different components of the Fourier transform for our moonlit sky observations are presented in Figure~\ref{fig:fc_b} for $B$ and \ref{fig:fc_I} for $I$. We have corrected all $Q_k$ and $U_k$ components for the median value of the map. We can clearly see at $k=2$ the same pattern for Q and U seen in Figure~\ref{fig:QU}, indicating as expected that this component carries the signal. All other harmonics have much lower contribution. The $k=0$ component, whose deviations from zero normally indicate anomalies with the WP, is reminiscent of the flat of section~\ref{sec:flat}. In fact, if we apply a flat correction prior to the Fourier analysis, the observed $Q_0$ pattern almost disappears entirely going below $Q_0 < 0.003$\%.  This demonstrates that although the final polarization degree is not significantly affected by a secondary flat correction, its minor contribution is clearly seen in the Fourier analysis. The $k=1$ and $k=3$ harmonics are quite different depending on wavelength: for $B$ we find a contribution that rises up to 0.2\% at the edges of the field, whereas in $I$ no clear pattern is seen and the signal is below 0.05\%. We attribute this partly to the pleochroic effect of the HWP that depends on wavelength. We also show the $P_k$ maps in Figure~\ref{fig:p_spec}. 

%%% Figure with Fourier Power spectrum 

From the secondary harmonics ($k\neq N/4$), we may also infer the error contribution, as follows:

\begin{eqnarray}
\Delta P &=& \frac{1}{N/2-1} \sum_{k=0;\,k\neq N/4}^{N/2} P_k, \\
\Delta \chi &=& \frac{1}{2}\arctan{\frac{\Delta P}{P}}. \nonumber    
\end{eqnarray}

\begin{figure}[h!t]
\centering
\includegraphics[trim={0.7cm 0cm 0.7cm 0.5cm},
clip,width=0.49\textwidth]
{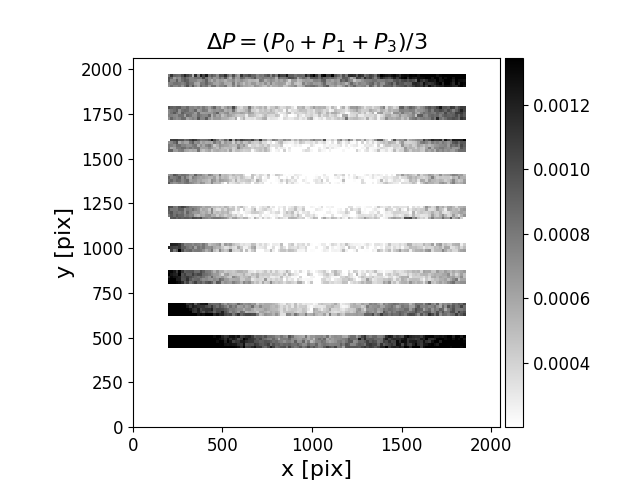} 
\includegraphics[trim={0.5cm 0cm 0.5cm 0.5cm },
clip,width=0.49\textwidth]
{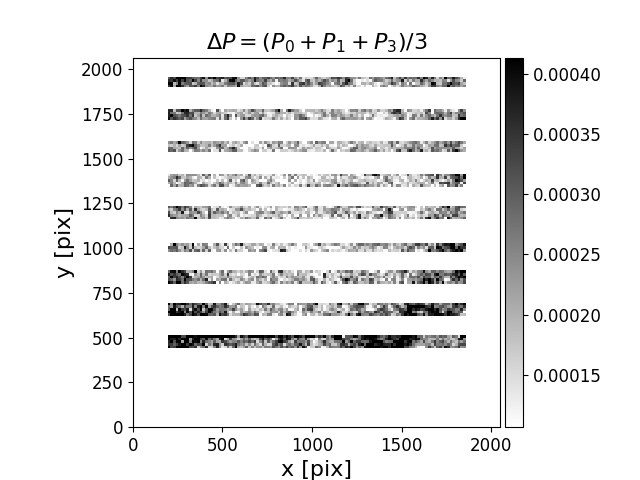}
\caption{ Binned (30$\times$30 pixel) sum of the Fourier harmonics except $k=2$ representing thus a characteristic error, $\Delta P$, for a moonlit blank sky field in $B$-band (top) and $I$-band (bottom). This is \emph{after} flat correction and median background $Q_{k,B},U_{k,B}$ correction.}\label{fig:p_err}
\end{figure}

In Figure~\ref{fig:p_err} we can see that the error budget on the polarization degree is less than 0.2\% in $B$ and less than 0.05\% in $I$. The median value and median absolute deviation of the error maps are shown in Table~\ref{table:error}. Thus the estimate based on the S/N (eq.~\ref{eq:polerr}), e.g. Figure~\ref{fig:polmap}, is larger and a conservative upper limit. We also confirm that the error is clearly larger than any of the residuals of the various models to correct the instrumental polarization (e.g. Table ~\ref{table:polparab}).

\begin{table}[h!t]
\centering
 \caption{Median and median absolute deviation (MAD) of the error maps obtained from the Fourier coefficients}\label{table:error}
 \scalebox{0.8}{
 \begin{tabular}{c|cccc}
    \hline
\hline  
Filter & $B$-band & $V$-band & $R$-band & $I$-band\\
\hline
<$\Delta Q [10^{-2}$\%]> & -0.213$\pm$2.271 & -0.055$\pm$2.760 & -0.066$\pm$1.477 & 0.014$\pm$0.863 \\
<$\Delta U [10^{-2}$\%]> & -0.128$\pm$1.548 & 0.319$\pm$1.547 & 0.054$\pm$0.762 & -0.007$\pm$0.647 \\
<$\Delta P [10^{-2}$\%]> & 5.822$\pm$2.278 & 6.773$\pm$2.791 & 3.559$\pm$1.227 & 2.222$\pm$0.598 \\
<$\Delta \chi$ [\textdegree]> & 4.552$\pm$1.317 & 4.338$\pm$1.308 & 2.081$\pm$0.827 & 1.085$\pm$0.400 \\
    \end{tabular}
    }
\end{table}